\documentclass[aip,amsmath,amssymb,article]{revtex4-1}

\usepackage{graphicx}
\usepackage{dcolumn}
\usepackage{bm}
\usepackage[mathlines]{lineno}

\usepackage[utf8]{inputenc}
\usepackage[T1]{fontenc}
\usepackage{mathptmx}
\usepackage{etoolbox}
\usepackage{ulem}
\usepackage{siunitx}

\usepackage{xcolor}
\usepackage{amsmath}
\usepackage{mathtools}
\usepackage{bm}
\usepackage{xcolor}

\usepackage[export]{adjustbox}

\usepackage{tikz}
\usetikzlibrary{arrows, positioning, calc}

\newcommand{\CV}{\mathcal{V}}
\newcommand{\kl}[2]{D_{\mathrm{KL}}(#1~\|~#2)}
\newcommand{\js}[2]{D_{\mathrm{JS}}(#1~\|~#2)}

\begin{document}
\title{Multi-scale Reconstruction of Turbulent Rotating Flows with Generative Diffusion Models}

\author{Tianyi Li}
\affiliation{Department of Physics and INFN, University of Rome ``Tor Vergata'', Via della Ricerca Scientifica 1, 00133 Rome, Italy}

\author{Alessandra S. Lanotte$^*$}
\email{alessandrasabina.lanotte@cnr.it}
\affiliation{Istituto di Nanotecnologia, CNR NANOTEC and INFN, Via per Monteroni, 73100 Lecce, Italy}

\author{Michele Buzzicotti}
\affiliation{Department of Physics and INFN, University of Rome ``Tor Vergata'', Via della Ricerca Scientifica 1, 00133 Rome, Italy}

\author{Fabio Bonaccorso}
\affiliation{Department of Physics and INFN, University of Rome ``Tor Vergata'', Via della Ricerca Scientifica 1, 00133 Rome, Italy}

\author{Luca Biferale} \affiliation{Department of Physics and INFN,
  University of Rome ``Tor Vergata'', Via della Ricerca Scientifica 1,
  00133 Rome, Italy}

\begin{abstract} We address the problem of data augmentation in a rotating
  turbulence set-up, a paradigmatic challenge in geophysical
  applications. The goal is to reconstruct information in
  two-dimensional (2D) cuts of the three-dimensional flow fields,
  imagining to have spatial gaps present within each 2D observed
  slice. We evaluate the effectiveness of different data-driven tools,
  based on diffusion models (DMs), a state-of-the-art generative
  machine learning protocol, and generative adversarial networks
  (GANs), previously considered as the best-performing method both in
  terms of point-wise reconstruction and the statistical properties of
  the inferred velocity fields. We focus on two different DMs recently
  proposed in the specialized literature: (i) RePaint, based on a
  heuristic strategy to guide an unconditional DM for flow generation
  by using partial measurements data and (ii) Palette, a conditional
  DM trained for the reconstruction task with paired measured and
  missing data. Systematic comparison shows that (i) DMs outperform
  the GAN in terms of the mean squared error and/or the statistical
  accuracy; (ii) Palette DM emerges as the most promising tool in
  terms of both point-wise and statistical metrics.  An important
  property of DMs is their capacity for probabilistic reconstructions,
  providing a range of predictions based on the same measurements,
  enabling for uncertainty quantification and risk assessment.
  \end{abstract}

\maketitle

\section{Introduction}
In atmospheric and oceanic forecasting, the accurate estimation of systems from incomplete observations is a challenging task \cite{le1986variational, bell2009godae, edwards2015regional, wang2021state, storer2022global}. These environments, often characterized by turbulent dynamics, require effective reconstruction techniques to overcome the common problem of temporally or spatially gappy measurements. The challenge arises from factors such as instrument sensitivity, the natural sparsity of observational data and also the absence of direct information, as e.g. in the case of deeper ocean layers \cite{shen2015missing, zhang2018missing, merchant2019satellite, wang2022gap, sammartino2020}. Established data assimilation techniques such as variational methods \cite{courtier1994strategy, yuan2023adjoint} and the ensemble Kalman filters \cite{houtekamer2001sequential, mons2021ensemble} effectively merge time-series observations with model dynamics to attack the inverse problem. When measurements are limited to a single time point, gappy proper orthogonal decomposition (POD) \cite{everson1995karhunen} and extended POD \cite{boree2003extended} deal with spatially incomplete data by exploiting pre-trained statistical relationships between measurements and missing information for the data augmentation goal. These POD-based methods are widely used in fluid mechanics \cite{venturi2004gappy, tinney2008low, discetti2019characterization} and geophysical fluid dynamics \cite{yildirim2009efficient, guemes2022super} to reconstruct flow fields.

POD-based methods are fundamentally linear yielding reconstructions with smooth flow properties, associated with few leading POD modes. In the context of turbulent flows, this implies that POD-like methods primarily emphasize large-scale structures \cite{li2023generative, li2023multi}. In recent years,  machine learning has led to an increasing number of successful applications in reconstruction tasks for simple and idealized fluid mechanics problems (see \cite{buzzicotti2023data} for a brief review). We mention super-resolution applications (i.e. finding high-resolution flow fields from low-resolution data) \cite{fukami2019super, liu2020deep, kim2021unsupervised}, inpainting (i.e. reconstructing flow fields having spatial damages) \cite{buzzicotti2021reconstruction, li2023multi}, and inferring volumetric flows from surface or two-dimensional (2D)-section measurements \cite{guastoni2021convolutional, matsuo2021supervised, yousif2023deep}. However, much remains to be clarified concerning benchmarks and challenges, 
and this is even more important for realistic turbulent set-up and at increasing flow complexity, e.g. for increasing Reynolds numbers. When dealing with turbulent systems, the quality of reconstruction tasks must be judged according to two different objectives: (i) the point-wise error, given by the succes to filling gappy or damaged regions of the instantaneous fields with data close to the ground truth configuration by configuration; (ii) statistical error, by reproducing statistical multi-scale and multi-point properties, such as probability distribution functions (PDFs), spectra, etc., of the 
system.

To move from proof-of-concept to quantitative benchmarks, in a previous work \cite{li2023multi}, we systematically compared POD-based methods with generative adversarial networks (GANs) \cite{goodfellow2014generative} using both point-wise and statistical reconstruction objectives for fully developed rotating turbulent flows, accounting for different gap sizes and geometries. GANs belong to the large family of generative models, i.e., machine learning algorithms that produce data according to a probability distribution optimized to  resembles that of the data used in the training. The learning task is made by two networks that compete with each
other: 
A first generative network is used to predict the data in the gap from the input measurement to obtain a good point-wise reconstruction; second, to overcome the lack of expressiveness in the multi-scale (with low energetic content) flow structures, a second adversarial network, called the discriminator, is used to optimize the statistical properties of the generated data. Contrary to expectations, despite their non-linearity,  GANs only matched the best linear POD techniques in point-wise reconstruction. However, GANs showed superior performance in capturing the statistical multi-scale non-Gaussian fluctuation characteristics of three-dimensional (3D) turbulent flow \cite{li2023multi}. 

From our previous comparative study, we also observed that GANs pose many challenges in the training processes, due to presence of instability and the necessity for hyper-parameters fine-tuning to achieve a suitable compromise in the multi-objective task. Furthermore, a common limitation of our GANs and POD-based methods is that they provide {\it only deterministic} reconstruction solution. This singular output contrasts with the intrinsic nature of turbulence reconstruction, which is a one-to-many problem with multiple plausible solutions. The ability to generate an ensemble of possible reconstructions is critical for practical atmospheric and oceanic forecasting, e.g., in relation to uncertainty quantification and risk assessment of rare, high-impact events \cite{smith2013uncertainty, hatanaka2023diffusion, asahi2023generating}.

More recently, diffusion models (DMs) \cite{ho2020denoising} have emerged as a powerful generative tool, showing exceptional success in domains such as computer vision \cite{ho2020denoising, nichol2021improved, dhariwal2021diffusion}, audio synthesis \cite{chen2020wavegrad}, and natural language processing \cite{brown2020language}, particularly outperforming GANs in image synthesis \cite{dhariwal2021diffusion}. Their applications have also extended to fluid dynamics for super-resolution \cite{shu2023physics}, flow prediction \cite{yang2023denoising} and Lagrangian trajectory generation \cite{li2023synthetic}. 
By introducing Markov chains to effectively generate data samples (see Section \ref{subsec:DM_flow_generation}), the implementation of DMs eliminates the need to resort to the less stable adversarial training of GANs, making DMs generally more stable in the training stage. Another characteristic of DMs is their inherent stochasticity in the generation process, which allows them to produce multiple outputs that adhere to the learned distribution conditioned on the same input.

This study focuses on the first attempt to using DMs for the reconstruction of 2D velocity fields of rotating turbulence, a complex system characterized by both large-scale vortices and highly non-Gaussian and intermittent small-scale fluctuations \cite{pouquet2013inverse,pouquet2017,alexakis2018cascades, buzzicotti2018energy, li2020flow}. Our objectives are twofold: first, we aim to make comprehensive comparisons with the best-performing GAN methods from our previous research, and second, we aim to investigate the effectiveness of DMs in probabilistic reconstruction tasks. The paper is organized as follows: in Section~\ref{sec:Methods}, we introduce the system under consideration and the two adopted strategies for flow reconstruction using DMs. The first is a heuristic conditioning method applied to an unconditional DM designed for flow generation, as demonstrated by RePaint \cite{lugmayr2022repaint}. The second strategy uses a supervised approach, training a DM conditioned on measurements, similar to the Palette method \cite{saharia2022image, saharia2022palette}.
In Section \ref{sec:comparison}, we discuss the performance of the two DMs in point-wise and statistical property reconstruction, in comparison with the previously analyzed GAN method \cite{li2023generative}.
In Section \ref{sec:ProbRecs}, we study the probabilistic reconstruction capacity of the DMs. 
We end with some comments in Section \ref{sec:conclusions}.

\section{Methods}\label{sec:Methods}
\subsection{Problem Setup and Data Preparation}
\begin{figure}
\begin{tikzpicture}
    \node[inner sep=0] (imgA) {
    \includegraphics[width=1.0\textwidth]{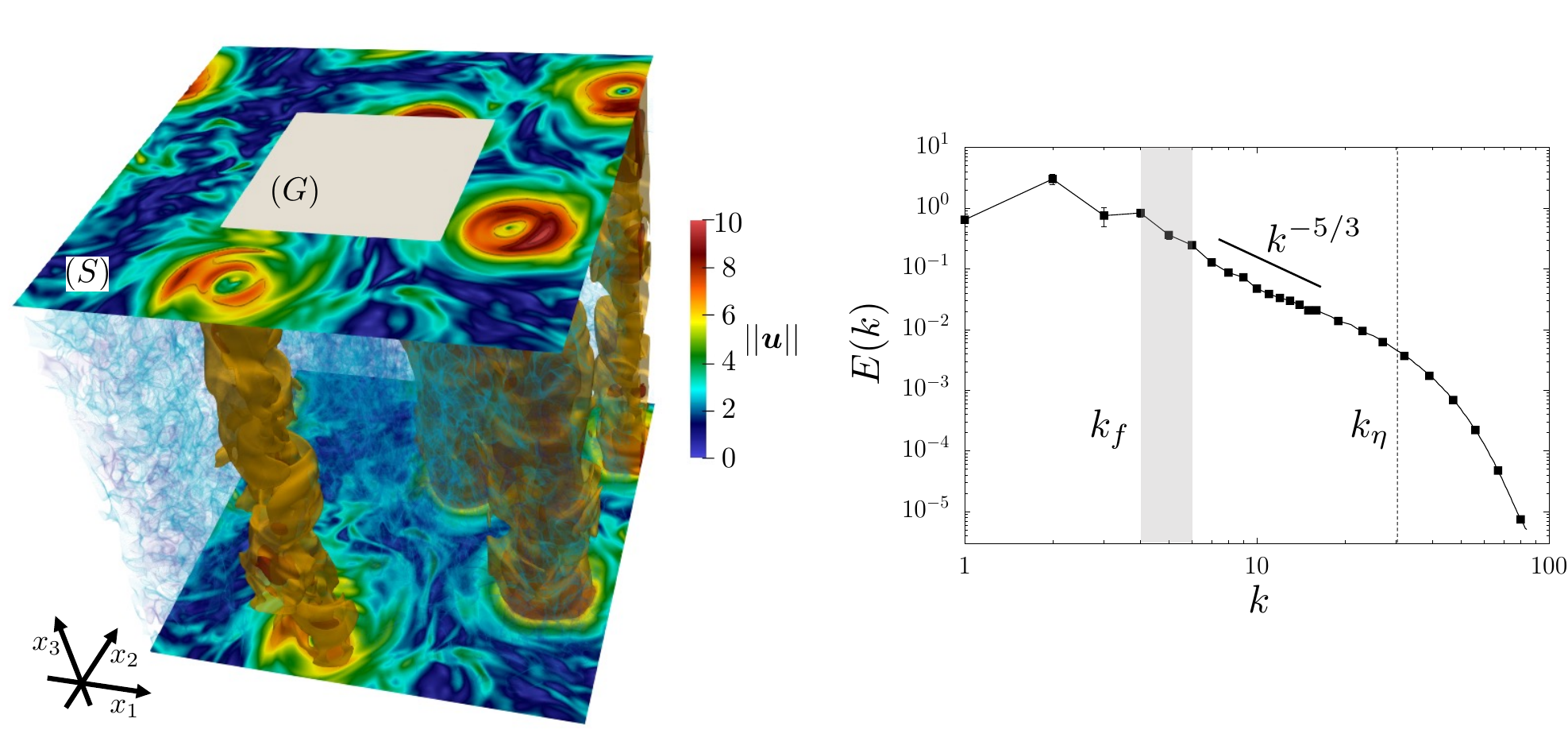}
    };
    \node[anchor=north west] at ([yshift=-1em]imgA.north west) {(a)};
    \node[anchor=north west] at ([xshift=0.5\textwidth+1em, yshift=-1em]imgA.north west) {(b)};
    \node[inner sep=0, below=1.5em of imgA] (imgB) {
    \includegraphics[width=1.0\textwidth]{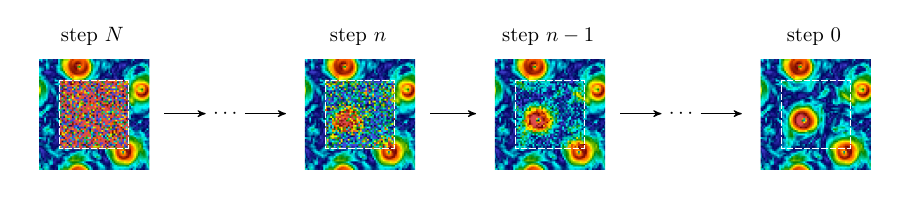}
    };
    \node[anchor=north west] at ([yshift=0.5em]imgB.north west) {(c)};
\end{tikzpicture}
    \caption{(\textbf{a}) Visualization of the velocity magnitude from
      a three-dimensional (3D) snapshot extracted from our numerical
      simulations. The two velocity planes (in the $x_1$-$x_2$
      directions) at the top and bottom of the integration domain show
      the velocity magnitude.  In the 3D volume we visualize a
      rendering of the small-scale velocity filaments developed by the
      3D dynamics. The gray square on the top level is an example of
      the damaged gap area, denoted as $(G)$, while the support where
      we suppose to have the measurements is denoted as $(S)$, and
      their union defines the full 2D image, $(I)=(S) \cup (G)$. A
      velocity contour around the most intense regions
      ($\|\bm{u}\|>6.35$) highlights the presence of the quasi-2D
      columnar structures (almost constant along $x_3$-axis), due to
      the effect of the Coriolis force induced by the frame
      rotation. (\textbf{b}) Energy spectra averaged over time. The
      range of scales where forcing is active is indicated by the gray
      band. The dashed vertical line denotes the Kolmogorov
      dissipative wavenumber. The reconstruction of the gappy area is
      based on a downsized image on a grid of $64^2$ collocation
      points, which corresponds to a resolution of the order of
      $1/k_\eta$. (\textbf{c}) Sketch illustration of the
      reconstruction protocol of a diffusion model (DM) in the
      backward phase (see later), which uses a Markov chain to
      progressively generate information through a neural
      network.}  \label{fig:3DVisual}
\end{figure}
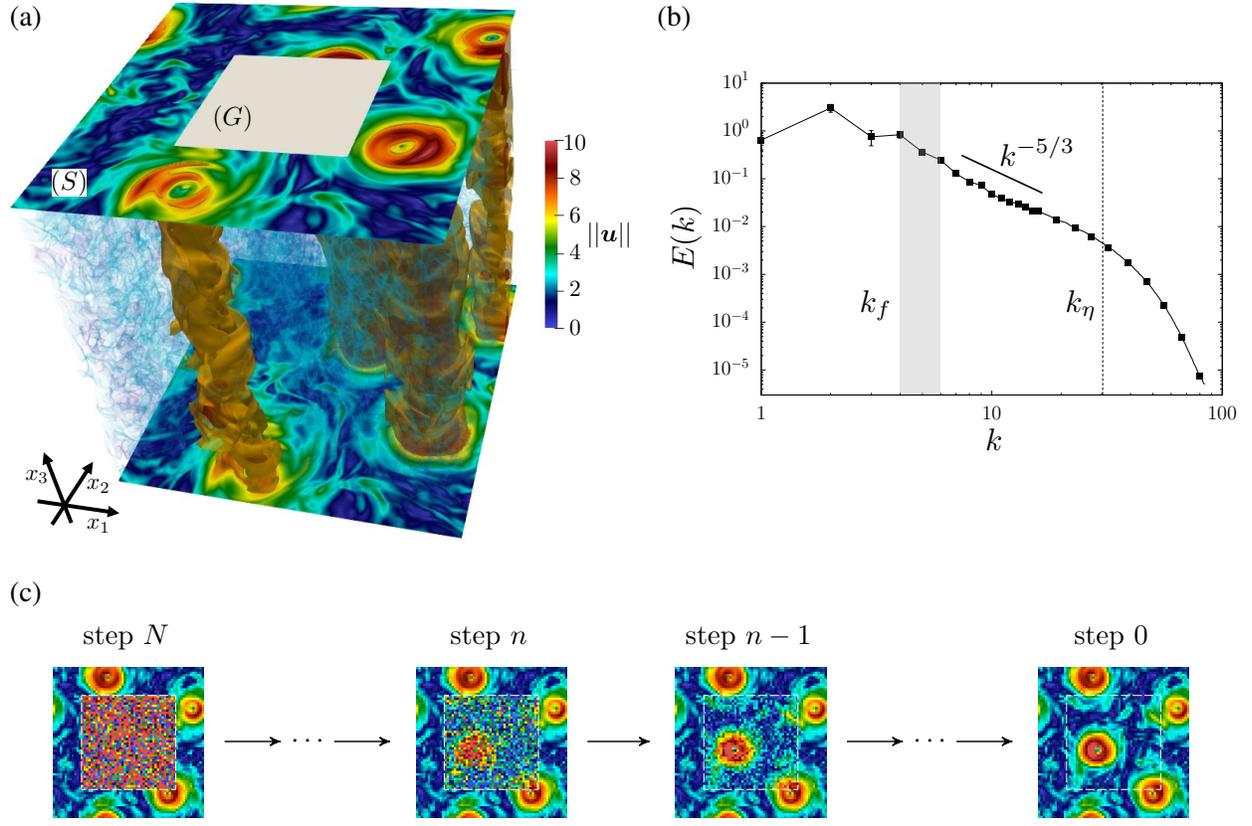

This study adopts the same experimental framework as our previous work \cite{li2023multi}, and explores possible improvements from DMs. We setup a mock field-measurement imagining to be able to obtain data from a gappy 2D slice of the original 3D volume  of rotating turbulence, orthogonal to the axis of rotation. The full 2D image is denoted as $(I)$, the support of the measured domain as $(S)$, and the support of the gap where we miss the data as $(G)$. Here $(G)$ represents a centrally located square gap of variable size, as shown in Figure \ref{fig:3DVisual}\textbf{a}. We use the TURB-Rot database \cite{biferale2020turb} obtained from direct numerical simulation (DNS) of the incompressible Navier-Stokes equations for rotating fluid in a 3D periodic domain, which can be written as
\begin{linenomath}
    \begin{equation}
    \label{eq:NS}
        \frac{\partial\bm{u}}{\partial t}+\bm{u}\cdot\bm{\nabla u}+2\bm{\varOmega}\times\bm{u}=-\frac{1}{\rho}\bm{\nabla}\tilde{p}+\nu\Delta\bm{u}+\bm{f},
    \end{equation}
\end{linenomath}
where $\bm{u}$ is the incompressible velocity, $\bm{\varOmega}=\varOmega\hat{\bm{x}}_3$ is the rotation vector, and $\tilde{p}$ represents the pressure modified by a centrifugal term. The regular, cubic grid has $N^3=256^3$ points. 
The statistically homogeneous and isotropic forcing $\bm{f}$ acts at
large scales around $k_f=4$, and it is the solution of a second-order
Ornstein–Uhlenbeck process \cite{sawford1991reynolds,
  buzzicotti2016lagrangian}. In the stationary state, with
$\varOmega=8$ the Rossby number is $Ro=\sqrt{\mathcal{E}}/(\varOmega
k_f)\approx0.1$, where $\mathcal{E}$ represents the kinetic
energy. The viscous dissipation $\nu\Delta \bm{u}$ is replaced by a
hyperviscous term $\nu_h\Delta^2\bm{u}$ to increase the inertial
range, while a large-scale linear friction term
$\alpha\Delta^{-1}\bm{u}$ is added to the r.h.s. of Equation
(\ref{eq:NS}) to reduce the formation of a large-scale condensate
\cite{alexakis2018cascades}, associated to the inverse energy cascade
well developed at this Rossby number. The Kolmogorov dissipative
wavenumber, $k_\eta=32$, is chosen as the scale at which the energy
spectrum begins to decay exponentially. An effective Reynolds number
is defined as $Re_{eff}=(k_0/k_\eta)^{-3/4}\approx13$, with the
smallest wavenumber $k_0=1$. The integral length scale is
$L=\mathcal{E}/\int kE(k)\,\mathrm{d}k\approx0.15L_0$, where
$L_0=2\pi$ is the domain length, and the integral time scale is
$T_L=L_0/{\mathcal{E}}^{1/2}\approx0.185$. For further details of DNS,
see \cite{biferale2020turb}, a sketch of the original 2D spectrum is
also shown in Figure \ref{fig:3DVisual}\textbf{b}.

Data were extracted from the DNS by sampling the full 3D velocity
field (Figure \ref{fig:3DVisual}\textbf{a}) during the stationary
stage at intervals of $\Delta t_s=5.41T_L$ to reduce temporal
correlation. We collected 600 early snapshots for training and 160
later snapshots for testing, with the two collections separated by
over $3400 T_L$ to ensure independence. To manage the data volume
while preserving complexity, the resolution of the sampled fields was
reduced from $256^3$ to $64^3$ using Galerkin truncation in Fourier
space, with the truncation wavenumber set to $k_\eta$. We then
selected $x_1$-$x_2$ planes at different $x_3$-levels and augmented
them by random shifts with the periodic boundary conditions, resulting
in a train/test split of 84,480/20,480 samples.

For a baseline comparison, we use the best-performing GAN tailored for
this setup in \cite{li2023multi}, which showed point-wise error close
to the best POD-based method and good multi-scale statistical
properties. In our analyses, we focus only on the velocity magnitude,
$u(x_1,x_2)=\|\bm{u}(x_1,x_2)\|$. Shortly, the GAN framework consists
of two competing convolutional neural networks: the first network is a
generator, that transforms input measurements into predictions for the
missing or damaged data; the second is a discriminator, that works to
discriminate between generated data and real fields. {The training of
  the generator minimizes a loss function consisting of mean squared
  error (MSE) and an adversarial loss provided by the discriminator,
  optimizing point-wise accuracy and statistical fidelity,
  respectively.} A more detailed description of the GAN can be found
in \cite{li2023multi}.

\begin{figure}
    \centering
    \includegraphics[width=1.0\textwidth]{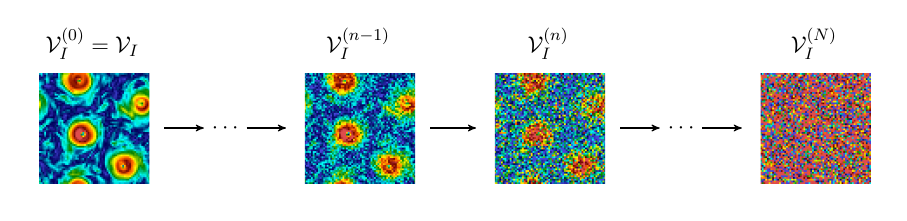}
    \caption{Diagram of the forward process in the DM framework. Starting with the original field $\CV_I^{(0)}=\CV_I$, Gaussian noise is incrementally added over $N$ diffusion steps, transforming the original $64^2$ image into white noise on the same resolution grid, $\CV_I^{(N)}$.}
    \label{fig:forward_turb}
\end{figure}

\subsection{DM Framework for Flow Field Generation}\label{subsec:DM_flow_generation}
Before moving to the more difficult task to inpaint a gap conditioned
on some partial measurements of each given image, we need to define
how to generate unconditional flow realizations.  Unlike GANs, which
map input noise to outputs in a single step, DMs use a Markov chain to
incrementally denoise and generate information through a neural
network, see Figure \ref{fig:3DVisual}\textbf{c} for a qualitative
visual example of one generation event. This finer-grained framework,
coupled with an explicit log-likelihood training objective, tends to
yield more stable training than the tailored loss functions of GANs,
but still has the capability of generating realistic samples. Another
feature of DMs is their inherent stochasticity in the generation
process, which allows them to produce multiple outputs that adhere to
the learned distribution conditioned on the same input.
\begin{figure}
\begin{tikzpicture}
    \node[draw, inner sep=0.5em] (imgB) {
    \begin{minipage}{0.98\textwidth}
        \includegraphics[valign=c, width=0.49\textwidth]{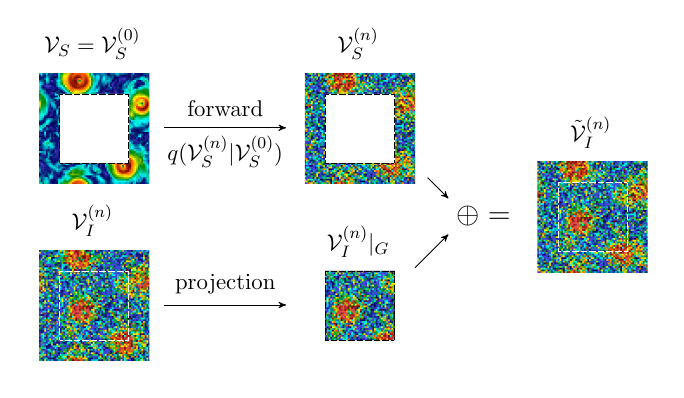}
        \includegraphics[valign=c, width=0.49\textwidth]{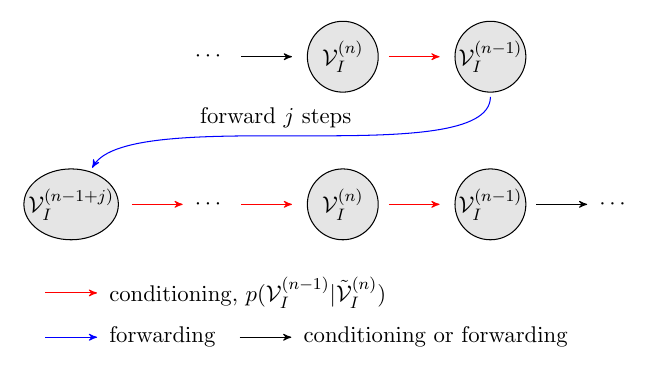}
    \end{minipage}};
    \node[anchor=north west] at (imgB.north west) {(c)};
    \node[anchor=north west, xshift=0.65\textwidth] at (imgB.north west) {(d)};
    \node[rectangle, draw, minimum width=1.\textwidth, inner sep=0, above=0.2cm of imgB] (imgA) {
    \begin{minipage}{0.755\textwidth}
        \centering
        \includegraphics[width=0.95\textwidth]{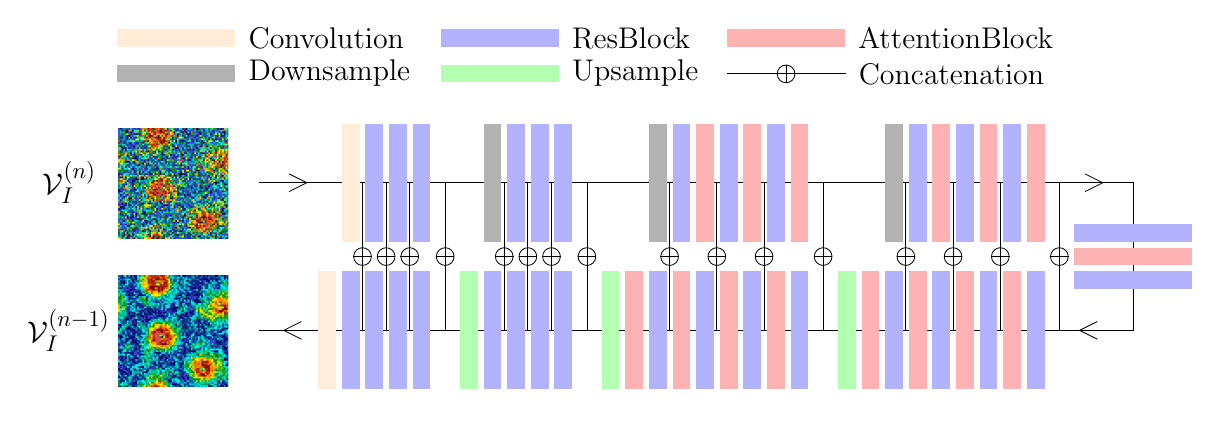}
        \par\vspace{0.07\textwidth}\par
        \includegraphics[width=0.95\textwidth]{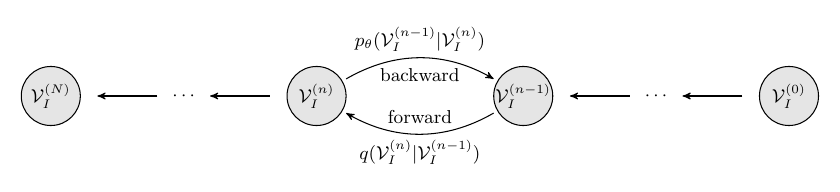}
    \end{minipage}};
    \node[anchor=north west] at (imgA.north west) {(a)};
    \node[anchor=north west] at ([yshift=-4cm] imgA.north west) {(b)};
    \node at ($(imgA.south west) + (0.175\textwidth, 0.165\textwidth)$) {
        \includegraphics[width=0.0884\textwidth]{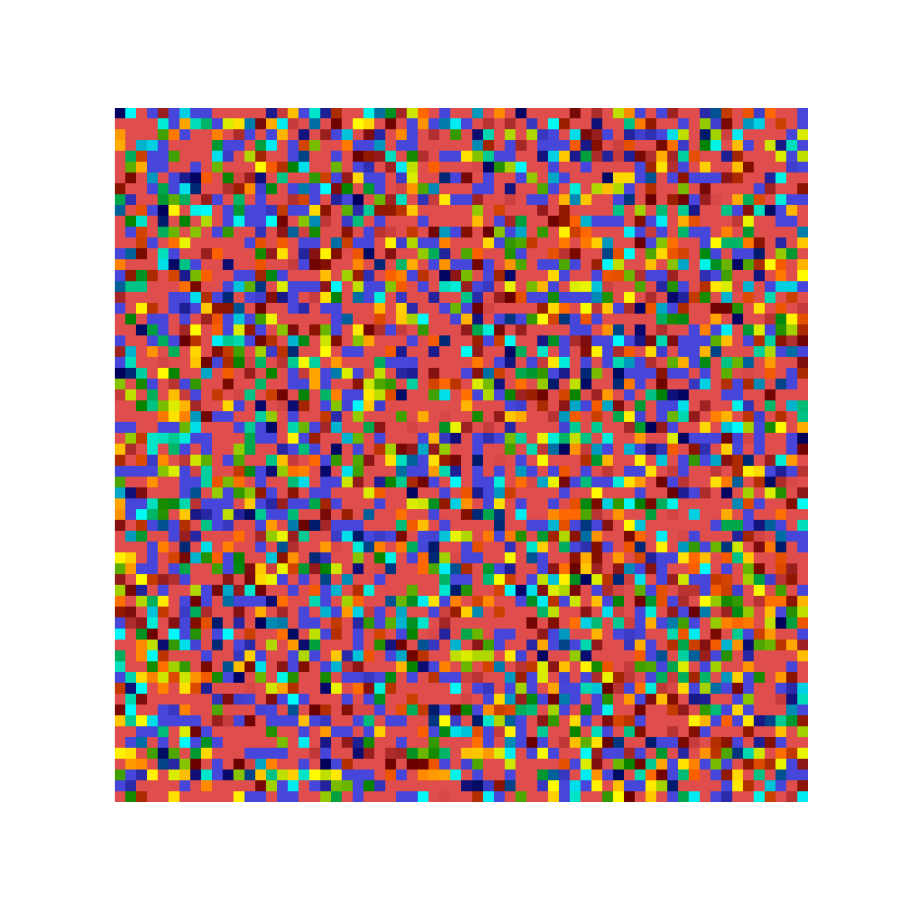}
    };
    \node at ($(imgA.south west) + (0.395\textwidth, 0.165\textwidth)$) {
        \includegraphics[width=0.0884\textwidth]{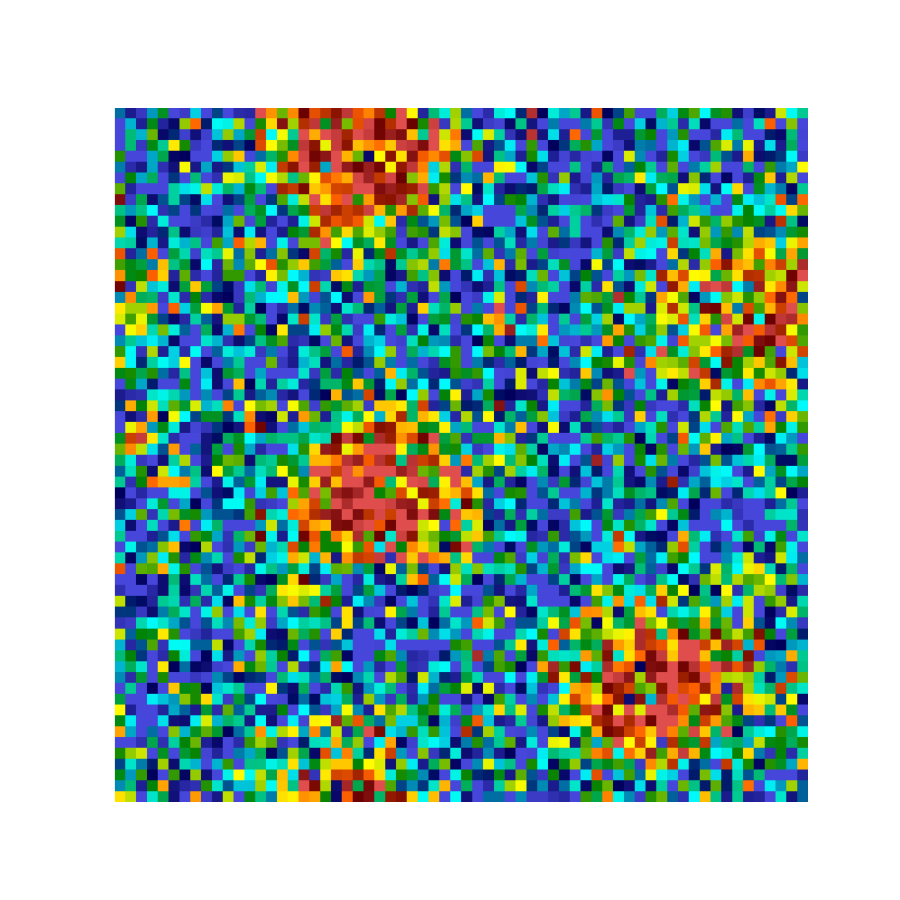}
    };
    \node at ($(imgA.south east) + (-0.4\textwidth, 0.165\textwidth)$) {
        \includegraphics[width=0.0884\textwidth]{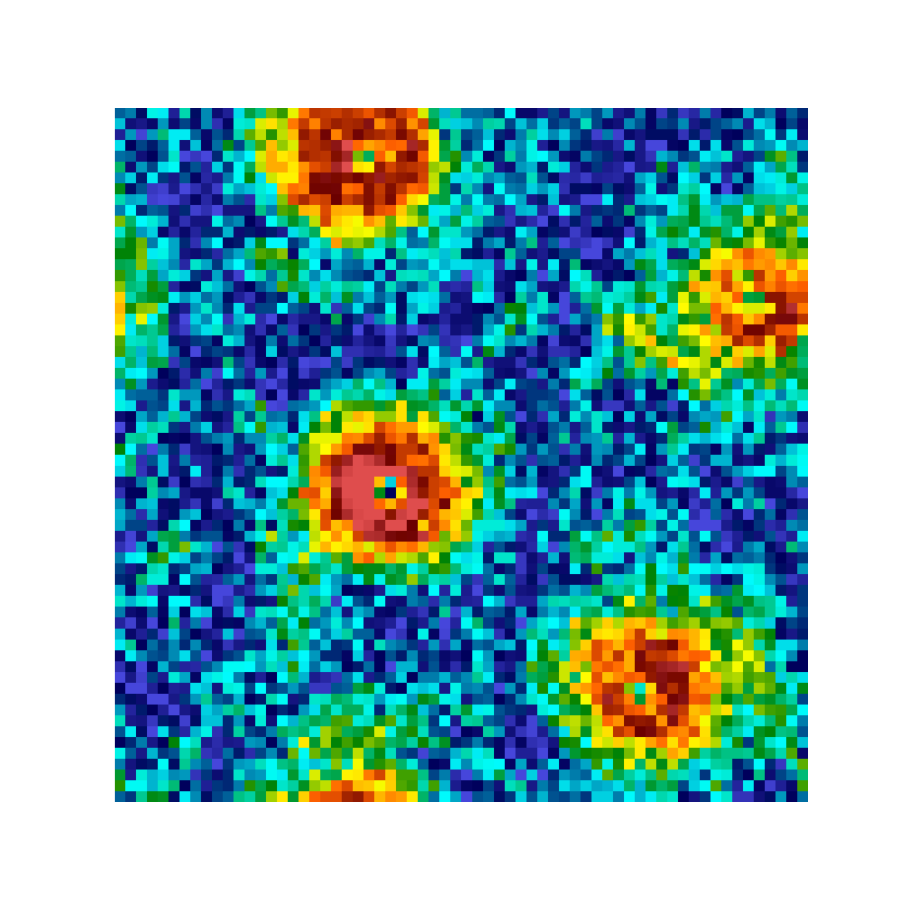}
    };
    \node at ($(imgA.south east) + (-0.18\textwidth, 0.165\textwidth)$) {
        \includegraphics[width=0.0884\textwidth]{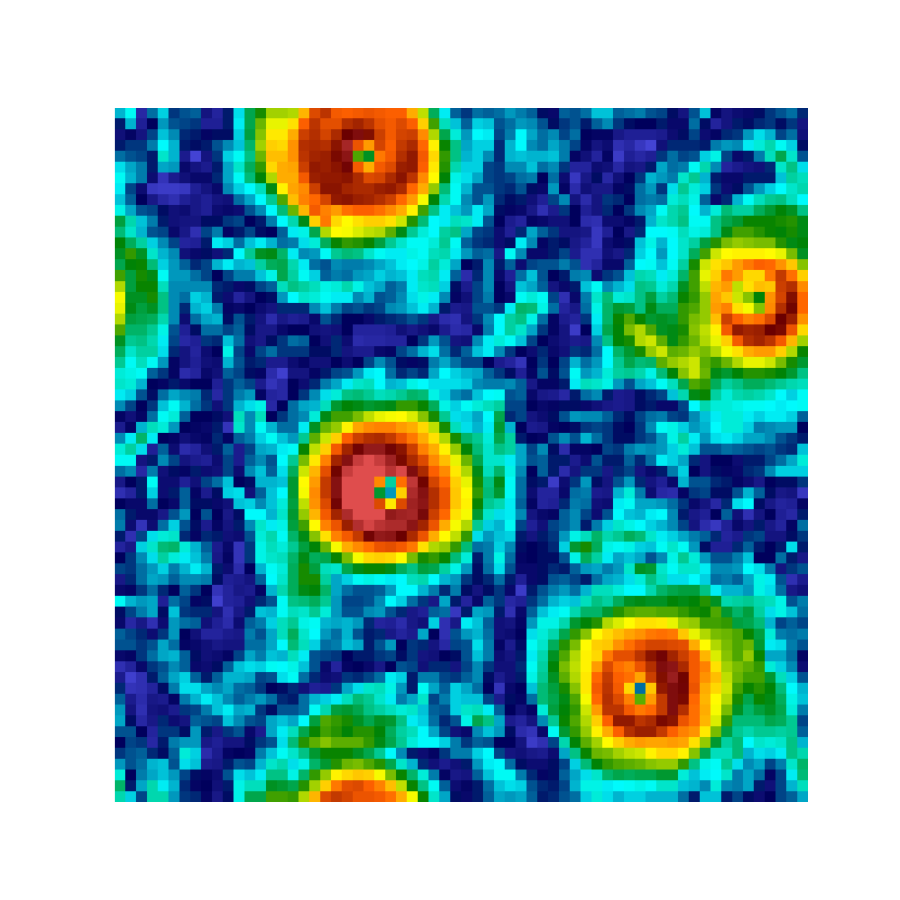}
    };
    \draw [->, >=latex, shorten >= 2pt, shorten <= 2pt, line width=2pt, color=red] ([yshift=2.4cm] imgA.south) -- ++(0, 1.4cm);
\end{tikzpicture}
\caption{Schematic representation of the DM flow field generation
  framework used by RePaint for flow reconstruction. Training stage:
  (\textbf{a}) the neural network architecture, U-Net
  \cite{ronneberger2015u}, that takes a noisy flow field as input at
  step $n$ and predicts a denoised field at step $n-1$; (\textbf{b})
  The scheme of the forward and backward diffusion Markov
  processes. The forward process (from right to left) incrementally
  adds noise over $N$ steps, while the backward process (from left to
  right), modeled by the U-Net, iteratively reconstructs the flow
  field by denoising the noisy data. More details on the network
  architecture can be found in Appendix \ref{sec:appB}. \\(\textbf{c,
    d}) Reconstruction stage starting from {\it damaged} fields with a
  square mask of variable size. (\textbf{c}) Conditioning the backward
  process with the measurement, $\CV_S$, involves projecting the noisy
  state of the entire 2D field, $\CV_I^{(n)}$, onto the gap region,
  $\CV_I^{(n)}|_G$, and combining it with the noisy measurement,
  $\CV_S^{(n)}$, obtained from the forward process up to the
  corresponding step. In this way we obtain $\tilde \CV_I^{(n)}$ and
  then enforce the conditioning by inputting it into a backward step.
  (\textbf{d}) A resampling approach is used to deal with the
  incoherence in $\tilde \CV_I^{(n)}$ introduced by the `rigid'
  concatenation. First, we peform a backward step to get $
  \CV_I^{(n-1)}$, and then some noise is added by $j$ forward steps
  (blue arrow). Finally, the field is resampled backwards by the same
  number of iterations, going back to the original
  step.\label{fig:RePaint}}
\end{figure}
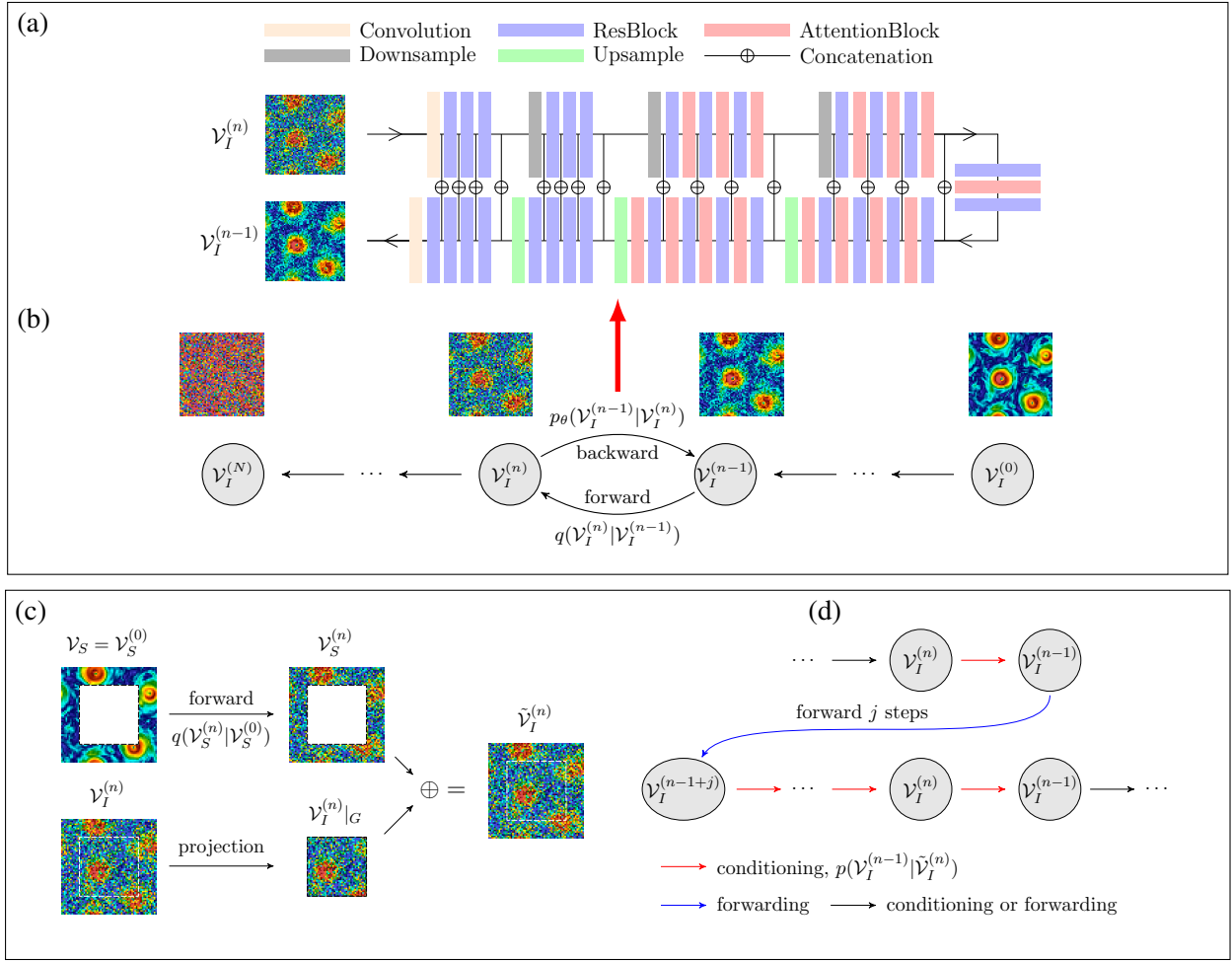
In this section, we introduce the DM framework for flow field
generation. The velocity magnitude field on the full 2D domain $(I)$
is denoted by $\CV_I=\{u(\bm{x})|\bm{x}\in I\}$, and the distribution
of this field is represented as $p(\CV_I)$. In order to train the
model we need first to produce a set of images with larger and larger
noise. To do that, the DM framework defines a \textit{forward process}
or \textit{diffusion process} that incrementally adds Gaussian noise
to the data until it becomes indistinguishable from white noise after
$N$ diffusion steps (Figure \ref{fig:forward_turb}).\\ This set of
{\it diffused} images is used for training a network to perform a {\it
  backward} denoising process, starting from the set of pure
i.i.d. Gaussian-noise 2D realizations and trying to reproduce the set
of images in the training data-set. Once accomplished the training,
one freezes the parameters of the network and uses it to generate
brand new images by sampling from any realization of pure random
images in the input (see Figure \ref{fig:RePaint}\textbf{a} for a
sketch summary). The forward diffusion process is expressed in terms
of a sequence of $N$ steps, conditioned on the original set of images,
i.e. for each image in the training data set we produce $N$ noisy
copies with an increasing amount of diffusion:
\begin{linenomath}
    \begin{equation}
q\left(\CV_I^{(1:N)}|\CV_I^{(0)}\right)\coloneqq\prod_{n=1}^{N}q\left(\CV_I^{(n)}|\CV_I^{(n-1)}\right),
    \end{equation}
\end{linenomath}
where $\CV_I^{(0)}=\CV_I$ is the initial magnitude field and $\CV_I^{(N)}\sim\mathcal{N}(\bm{0},\bm{I})$ represents the final white-noise state, an ensemble of Gaussian images made of uncorrelated pixels  with zero mean and unit variance. The notation $\CV_I^{(1:N)}$ is used to denote the entire sequence of generated noisy fields, $\CV_I^{(1)},\ldots,\CV_I^{(N)}$. \\Each step, $n = 1,\dots, N$,  of the forward process can be directly obtained as
\begin{linenomath}
    \begin{equation}\label{equ:forward}
        q\left(\CV_I^{(n)}|\CV_I^{(n-1)}\right)\to\CV_I^{(n)}\sim\mathcal{N}\left(\sqrt{1-\beta_n}\CV_I^{(n-1)},\beta_n\bm{I}\right),
    \end{equation}
\end{linenomath}
which implies sampling from a Gaussian distribution where the mean of
$\CV_I^{(n)}$ is given by $\sqrt{1-\beta_n}\CV_I^{(n-1)}$ and the
variance is $\beta_n\bm{I}$. The variance schedule
$\beta_1,\ldots,\beta_N$ is predefined to allow a continuous
transition to the pure Gaussian state. For more details on the
variance schedule and other aspects of the DMs used in this study, see
Appendix \ref{sec:appB}.

The DM trains a neural network to approximate the reverse process of
Equation (\ref{equ:forward}), denoted as $p_\theta
\left(\CV_I^{(n-1)}|\CV_I^{(n)}\right)$. This approximation allows the
generation of new velocity magnitude fields from Gaussian noise,
$p\left(\CV_I^{(N)}\right)=\mathcal{N}\left(\bm{0},\bm{I}\right)$,
through a \textit{backward process} (see Figure
\ref{fig:RePaint}\textbf{a}) described by
\begin{linenomath}
    \begin{equation}
        p_\theta\left(\CV_I^{(0:N)}\right)\coloneqq p\left(\CV_I^{(N)}\right)\,\prod_{n=1}^{N}p_\theta\left(\CV_I^{(n-1)}|\CV_I^{(n)}\right).
    \end{equation}
\end{linenomath}
{Where it is important to notice that the stochasticity in the process
  allows for the production of different final images even when
  starting from the same noise.}  In the continuous diffusion limit,
characterized by sequences of small values of $\beta_n$, the backward
process has a functional form identical to that of the forward
process, as discussed in \cite{feller2015theory,
  sohl2015deep}. Consequently, the neural network is tasked with
predicting the mean $\mu_\theta(\CV_I^{(n)},n)$ and covariance
$\Sigma_\theta(\CV_I^{(n)},n)$ of a Gaussian distribution:
\begin{linenomath}
    \begin{equation}\label{equ:backward}
        p_\theta\left(\CV_I^{(n-1)}|\CV_I^{(n)}\right)\to\CV_I^{(n-1)}\sim\mathcal{N}\left(\mu_\theta(\CV_I^{(n)},n),\Sigma_\theta(\CV_I^{(n)},n)\right).
    \end{equation}
\end{linenomath}
The neural network is optimized to minimize an upper bound of the negative log likelihood, 
\begin{linenomath}
    \begin{equation}\label{equ:nll}
        \mathbb{E}_{q(\CV_I^{(0)})}[-\log(p_\theta(\CV_I^{(0)}))].
    \end{equation}
\end{linenomath}
{This training objective tends to result in more stable training compared to the tailored loss functions used in GANs.}
For a detailed derivation of the loss function and insights into the training details, please refer to Appendix \ref{sec:appA}.



\subsection{Flow Field Data Augmentation with DMs: RePaint and Palette Strategies}
{\sc RePaint}. The RePaint approach aims to reconstruct missing information in the flow field using a DM that has been trained to generate the full 2D flow field from Gaussian noise as described in the section above, without any conditioning on measured data, and without relying on any further model training. To achieve the correct reconstruction, RePaint aims to ensure the conditioning on the measurements only by redesigning an ad-hoc generation protocol~\cite{lugmayr2022repaint}.
As discussed above, during training DM learns to approximate the backward transition probability to step on a sample $\CV_I^{(n-1)}$, only from the knowledge of the sample obtained in the previous step $\CV_I^{(n)}$, hence, DM models the one-step backward transition probability, $p_\theta \left(\CV_I^{(n-1)}|\CV_I^{(n)} \right )$. 
The goal of RePaint is to set up a generative process where this backward probability is also conditioned on some measured data, denoted as $\CV_S$. In this way, each new sample in the backward direction is generated from the one-step backward conditioned probability, defined as $p_\theta\left(\CV_I^{(n-1)}|\CV_I^{(n)},\CV_S\right)$.
To achieve this goal, RePaint substitutes the DM model input, $\CV_I^{(n)}$, with another 2D field, $\tilde{\CV}_I^{(n)}$, which is given by the union of $\CV_I^{(n)}$ projected only to have support inside the gap $(G)$ and the measured data on the support $(S)$ propagated at step $n$ according to the forward process, namely, $\tilde{\CV}_I^{(n)}=\CV_I^{(n)}|_G\oplus\CV_S^{(n)}$.
In summary, at any generic backward step $n$, RePaint approximates the conditional backward probability as follows:
\begin{linenomath}
    \begin{equation}
        p_\theta\left(\CV_I^{(n-1)}|\CV_I^{(n)},\CV_S\right)\approx p_\theta\left(\CV_I^{(n-1)}|\tilde{\CV}_I^{(n)}\right)\quad\text{where}\quad\tilde{\CV}_I^{(n)}=\CV_I^{(n)}|_G\oplus\CV_S^{(n)}\,.
    \end{equation}
\end{linenomath}
Here, $\CV_I^{(n)}|_G$, represents the projection of the sample generated by the backward process at step $n$ projected inside the gap region (the central square), while, $\CV_S^{(n)}$, is the noisy version of the measured data (outside the square gap) that is obtained by a forward propagation up to step $n$ of the measurements. At this point, $\tilde{\CV}_I^{(n)}$, replacing $\CV_I^{(n)}$, is given as input to the model and it is used to obtain the next sample at step $n-1$, $\CV_I^{(n-1)}$, see Figure \ref{fig:RePaint}\textbf{c}. 

The propagation of information from the measurements into the gap happens, thanks to the application of the non-linear (and non-local) function approximated by the U-Net employed in the DM. Hence, the output of the U-Net, describing the probability of moving from step $n$ to $n-1$, is the result of non-local convolutions mixing information in the two regions $(S)$ and $(G)$.
In this way, the model mitigates the discontinuities generated across the gap by merging the generated and the measured data. Furthermore, to allow a deeper propagation of information, improving correlations between the measurements and the generated data, RePaint employs a resampling strategy \cite{lugmayr2022repaint}. 
The idea of resampling, as shown schematically in Figure \ref{fig:RePaint}\textbf{d}, is that each sample at step $n-1$, extracted from conditioned probability, $p_\theta\left(\CV_I^{(n-1)}|\tilde{\CV}_I^{(n)}\right)$, is not directly used as input to move backward at step $n-2$, but instead it is first propagated forward for $j$ steps (by adding more noise) before returning according to the conditioned backward process at step $n-1$. 
This operation gives the U-Net model the opportunity to iterate the propagation of information from the measured region inside the gap. Resampling can be applied at different steps multiple times, resulting in a back-and-forth progression during the generation process, as opposed to a monotonic backward progression from $n=N$ to $n=0$. Further details, such as the network architecture and other parameters, can be found in Appendix \ref{sec:appB}.
As demonstrated in computer vision applications \cite{richardson2021encoding, chung2022diffusion, lugmayr2022repaint, zhang2023towards}, this strategy has the advantage of being easily generalizable to diverse tasks, such as free-form inpainting with arbitrary mask shapes. However, this introduces several new challenges in the design of such a convoluted generation protocol, which is neither trivial in its optimization nor in its implementation.
\\

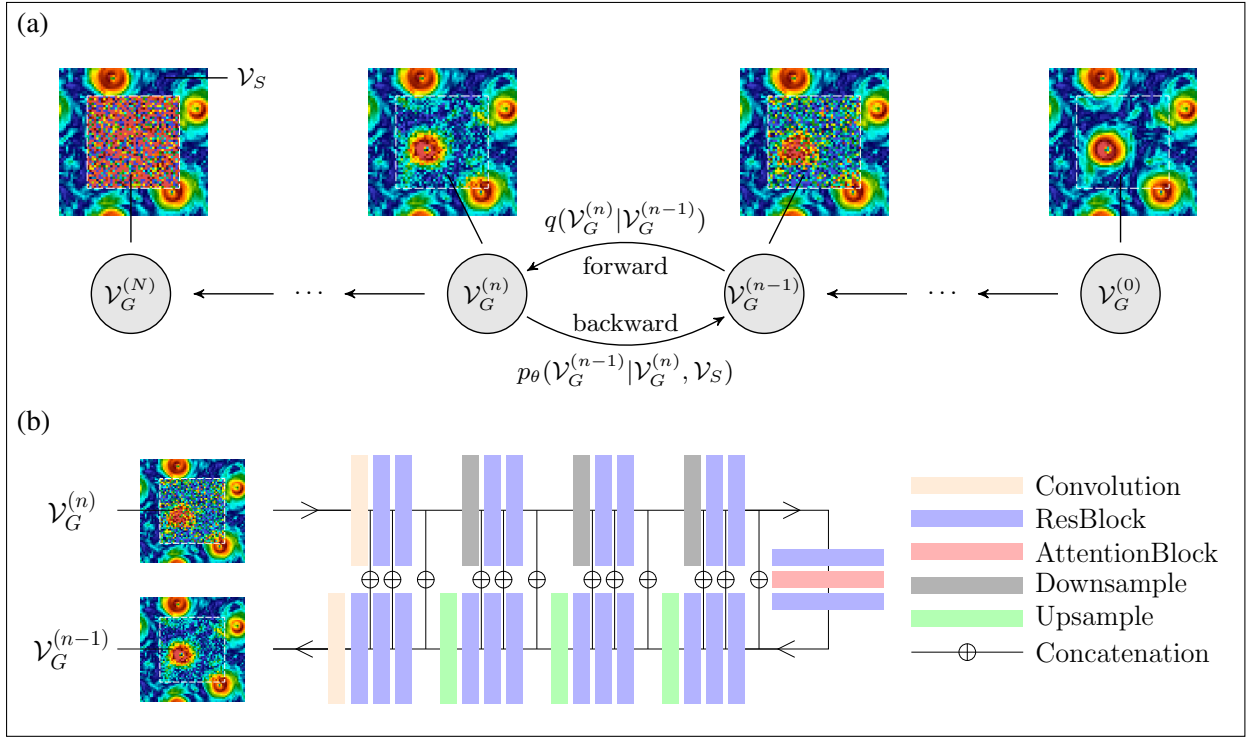
\begin{figure}
\begin{tikzpicture}
    \node[inner sep=0] (imgA) {
    \includegraphics[width=1.0\textwidth]{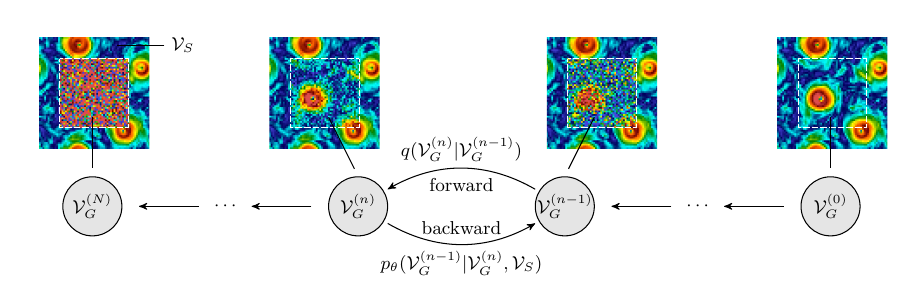}};
    \node[anchor=north west] at ([yshift=0.5em]imgA.north west) {(a)};
    \node[inner sep=0, below=0cm of imgA] (imgB) {
    \includegraphics[width=1.0\textwidth]{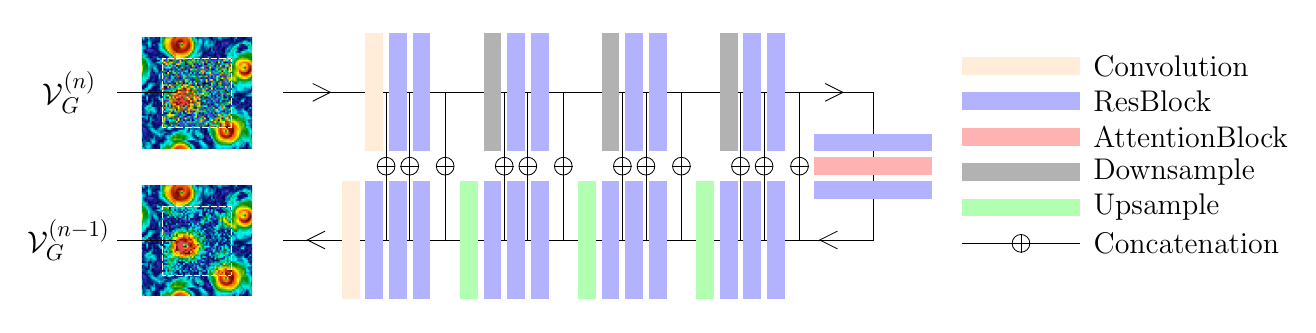}};
    \node[anchor=north west] at ([yshift=0.8em]imgB.north west) {(b)};
    \draw ([yshift=0.5em]imgA.north west) rectangle (imgB.south east);
\end{tikzpicture}
\caption{Schematic of the DM Palette protocol. (\textbf{a}) In the backward process (from left to right), we start from pure noise in the gap, $\CV_G^{(N)}$, combined with the measurements in the frame, $\CV_S$, to progressively denoise the missing information using the U-Net architecture described in panel \textbf{b}. (\textbf{b}) A sketch of the U-Net integrating the measurement, $\CV_S$, and the noisy data within the gap, $\CV_G^{(n)}$, for a backward step.\label{fig:Palette}}
\end{figure}

{\sc Palette.} An alternative approach to perform flow field reconstruction is to train the DM directly to learn the backward probability distribution conditioned on the measured data, previously introduced as $p_\theta\left(\CV_G^{(n-1)}|\CV_G^{(n)},\CV_S\right)$. 
This method, called Palette, has been successfully used in various computer vision applications such as image-to-image translation tasks \cite{saharia2022image, saharia2022palette}. 
The idea is to train a U-Net using the same strategy as any unconditioned DM, but giving the network as input the additional information coming from the measurements, at any step during the diffusion process. This allows the model to learn during training how to use information from available data to achieve optimal reconstruction inside the gap. In addition, unlike the RePaint method, Palette always uses the measured data without adding noise. In this way, the forward process can be defined as for the pure generation case, but it takes place only within the gap region, while the data on the support, $\CV_S$, are frozen throughout the diffusion process and serves as an additional input to the model. 
A schematic summary of the Palette approach is shown in Figure \ref{fig:Palette}.
Once the DM model is trained, since the reconstruction process is Markovian as in the standard generative DM, the conditional probability of the reconstructed field, $p_\theta\left(\CV_G^{(0)}|\CV_S\right)$, can be determined through the following iterative process:
\begin{linenomath}
    \begin{equation}        p_\theta\left(\CV_G^{(0)}|\CV_S\right)=p\left(\CV_G^{(N)}\right)\,\prod_{n=1}^{N}p_\theta\left(\CV_G^{(n-1)}|\CV_G^{(n)},\CV_S\right),
    \end{equation}
\end{linenomath}
starting from any Gaussian noise $\CV_G^{(N)}$. 
To facilitate the comparison with the GAN model implemented in our previous works~\cite{buzzicotti2021reconstruction,li2023multi}, we trained a separate Palette model for each fixed mask size. Let us stress that both methods are capable of training on a free-form mask \cite{yu2019free}. More details on Palette are in Appendix \ref{sec:appB}.

\section{Comparative Analysis of DMs and the GAN in Flow Reconstruction}\label{sec:comparison}
To provide a systematic comparison between DMs and the GAN in flow reconstruction, we focus on cases where the 2D velocity magnitude fields have a central square gap of variable size, spanning $0.1< l/l_0 < 1$, where $l_0$ is the size of the whole flow domain. In this section, only one reconstruction realization is performed for all image data in the testing ensemble, i.e. we do not further explore the possibility of assessing the robustness of the prediction by sampling over the ensemble of predicted images (see next section). The initial evaluation focuses on the reconstructed single-point velocity magnitude, which is strongly influenced by large-scale coherent structures. Then, we analyze the reconstruction process from a multi-scale perspective, by examining the statistical properties of the gradient of the reconstructed velocity magnitude, and by looking at other scale-dependent statistics in both real and Fourier space.

\subsection{Large-scale Information}
To quantify the reconstruction error between the predicted velocity magnitude, $u_G^{(p)}$, and the true velocity magnitude, $u_G^{(t)}$, within the gap region, we introduce the normalized MSE as follows:
\begin{linenomath}
    \begin{equation}\label{equ:MSE}
        \mathrm{MSE}(u_G)=\langle\Delta_{u_G}\rangle/E_{u_G}\,.
    \end{equation}
\end{linenomath}
Here $\Delta_{u_G}$ represents the spatially averaged $L_2$ error in the central, gappy region for a single flow configuration, and it is calculated as
\begin{linenomath}
    \begin{equation}
        \Delta_{u_G}=\frac{1}{A_G}\int_G|u_G^{(p)}(\bm{x})-u_G^{(t)}(\bm{x})|^2\mathrm{d}\bm{x}\,,
    \end{equation}
\end{linenomath}
where $A_G$ denotes the area of the gap. Averaging $\langle\cdot\rangle$ is done over the test data set. The normalization factor, $E_{u_G}$, is defined as the product of the standard deviations of the predicted and true velocity magnitudes within the gap:
\begin{linenomath}
    \begin{equation}
        E_{u_G}=\sigma_G^{(p)}\sigma_G^{(t)},
    \end{equation}
\end{linenomath}
where
\begin{linenomath}
    \begin{equation}
        \sigma_G^{(p)}=\frac{1}{A_G^{1/2}}\int_G \langle (u_G^{(p)})^2\rangle^{1/2}\mathrm{d}\bm{x}
    \end{equation}
\end{linenomath}
and $\sigma_G^{(t)}$ is similarly defined. {This choice for the normalization term, $E_{u_G}$, ensures that predictions with significantly low or high energy levels will result in a large MSE.}

In our analysis, we use the Jensen-Shannon (JS) divergence to assess the distance between the PDF of a predicted quantity and the PDF of the true data. Specifically, the JS divergence applied to two distributions $P(x)$ and $Q(x)$ defined on the same sample space is 
\begin{linenomath}
    \begin{equation}
        \js{P}{Q}=\frac{1}{2}\kl{P}{M}+\frac{1}{2}\kl{Q}{M},
    \end{equation}
\end{linenomath}
where $M=\frac{1}{2}(P+Q)$ and
\begin{linenomath}
    \begin{equation}
        \kl{P}{Q}\equiv\int_{-\infty}^{\infty}P(x)\log\left(\frac{P(x)}{Q(x)}\right)\mathrm{d}x
    \end{equation}
\end{linenomath}
is the Kullback-Leibler (KL) divergence. As the two distributions get closer, the value of the JS divergence  becomes smaller, with a value of zero indicating that $P$ and $Q$ are identical. 

\begin{figure}
	\includegraphics[width=1.0\textwidth]{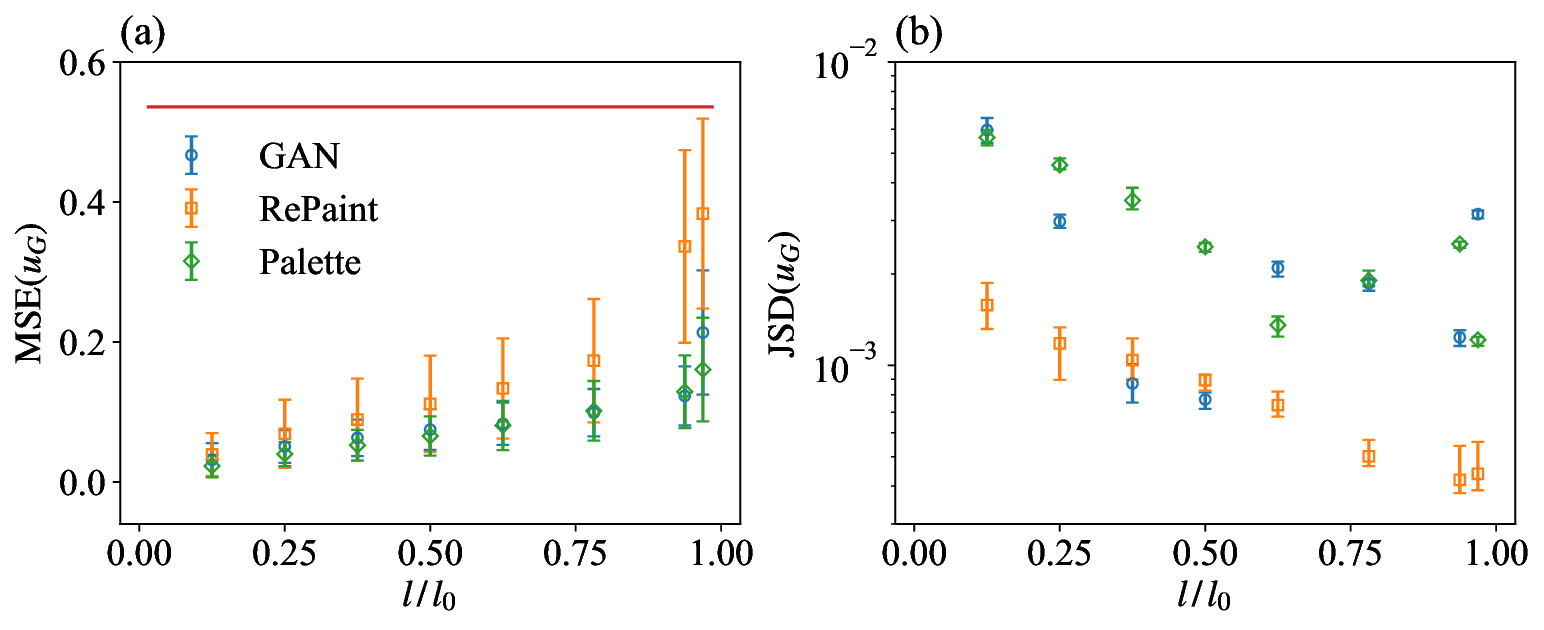}
    \caption{(\textbf{a}) The mean squared error (MSE) between the true and the generated velocity magnitude, as obtained from GAN, RePaint and Palette, for a square gap with variable size. Error bars indicate the standard deviation. The red horizontal line represents the uncorrelated baseline MSE, $\approx 0.54$. (\textbf{b}) The Jensen-Shannon (JS) divergence between the probability density functions (PDFs) for the true and generated velocity magnitude. The mean and error bars represent the average and range of variation of the JS divergence across 10 batches, each with 2048 samples.}  \label{fig:MSE_JSD}
\end{figure}

\begin{figure}
    \centering
    \includegraphics[width=\textwidth]{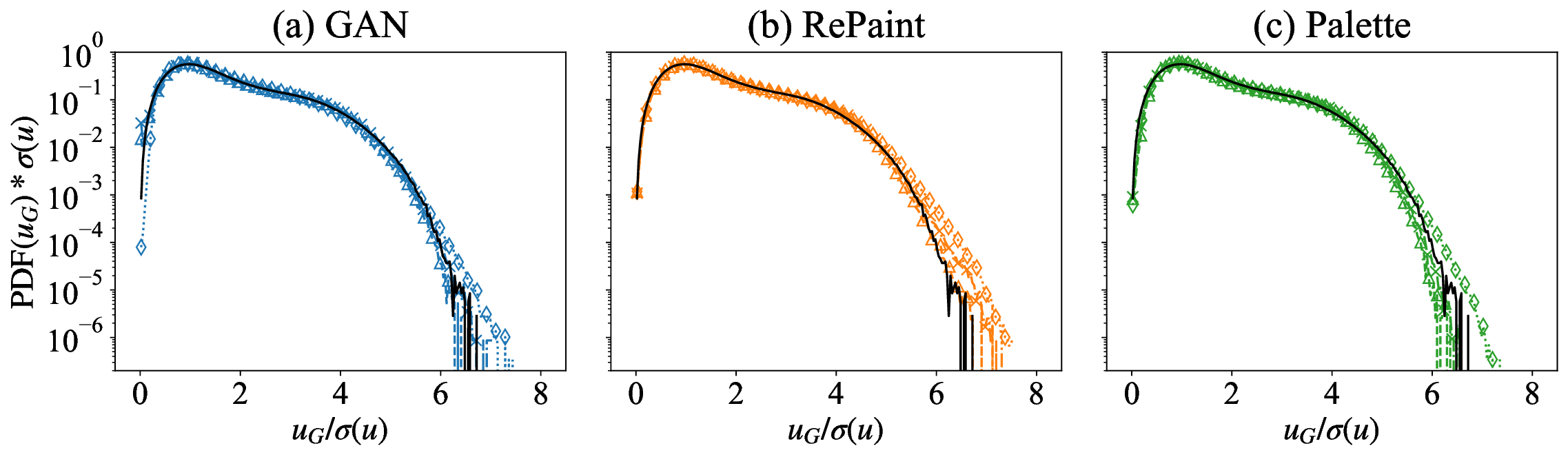}
\caption{PDFs of the velocity magnitude in the missing region obtained
  from (\textbf{a}) GAN, (\textbf{b}) RePaint and (\textbf{c}) Palette
  for a square gap of variable size $l/l_0=24/64$ (triangle), $40/64$
  (cross), and $62/64$ (diamond). The PDF of the true data over the
  whole region is plotted for reference (solid black line) and
  $\sigma(u)$ is the standard deviation of the original data over the
  full domain.\label{fig:PDF}}
\end{figure}

Figure \ref{fig:MSE_JSD}\textbf{a} shows the $\mathrm{MSE}(u_G)$ as a
function of the normalized gap size, $l/l_0$. It shows that Palette
achieves a comparable MSE with respect to GAN, for most gap
sizes. Only for the largest gap size, $l/l_0=62/64$, the MSE of
Palette is significantly better than that of GAN. On the other hand,
RePaint has a larger MSE for all sizes compared to the other two
methods, demonstrating the limitations of the RePaint approach in
enforcing correlations between measurements and generated data without
being specifically trained on a reconstruction problem as the other
two approaches.  The red baseline, derived from predictions using
randomly shuffled test data, represents the case where the predictions
guess the exact statistical properties, $\langle u_G^{(p)} \rangle =
\langle u_G^{(t)}\rangle$ and $\langle (u_G^{(p)})^2 \rangle = \langle
(u_G^{(t)})^2 \rangle$, but lose all correlation with the
measurements, $\langle u_G^{(p)}u_G^{(t)} \rangle= \langle u_G^{(p)}
\rangle \langle u_G^{(t)} \rangle$.
We now examine the velocity magnitude PDFs as predicted by the different methods and compare them with the true data. In Figure \ref{fig:MSE_JSD}\textbf{b} we present the JS divergence between the predicted and true velocity magnitudes, denoted as 
$\mathrm{JSD}(u_G)=\js{\mathrm{PDF}(u_G^{(p)})}{\mathrm{PDF}(u_G^{(t)})}$. 
First of all, it is important to highlight that all the $\mathrm{JSD}(u_G)$ values are well below $10^{-2}$, suggesting that there is always a close match between the PDFs of the reconstructed and that of the true velocity magnitude. 
The agreement between the different PDFs is also shown in Figure \ref{fig:PDF}, where one can see the extremely good performance of all models to closely match the PDFs of the generated velocity magnitude with the ground truth one. 
Going back to the results presented in Figure \ref{fig:MSE_JSD}\textbf{b}, it is possible to note that in the small gap region, $l/l_0 \le 0.4$, there is a monotonic behavior of the JS divergence, which tends to decrease as the gap increases. This can be interpreted by the fact that the main contribution to the JS divergence is due to statistical fluctuations in the PDF tails which are less accurately estimated when the gap is small. This behavior is clearly visible in the results of the RePaint approach, which shows a monotonic decrease in the JS divergence over the whole range of gaps analyzed. The same effect is not visible in the other approaches in the range above $l/l_0=0.5$. The reason is probably due to the fact that both GAN and Palette rely on different training to reconstruct different gap sizes, and the fluctuations due to the training convergence could be underestimated. The non-monotonicity is much more pronounced in the GAN results, as this approach is known to be less stable during training. 
The analysis shows that, contrary to the other two approaches, RePaint trained on the pure generation without any conditioning is the best method to obtain a statistical representation of the true data. 

\begin{figure}
    \centering
    \includegraphics[width=\textwidth]{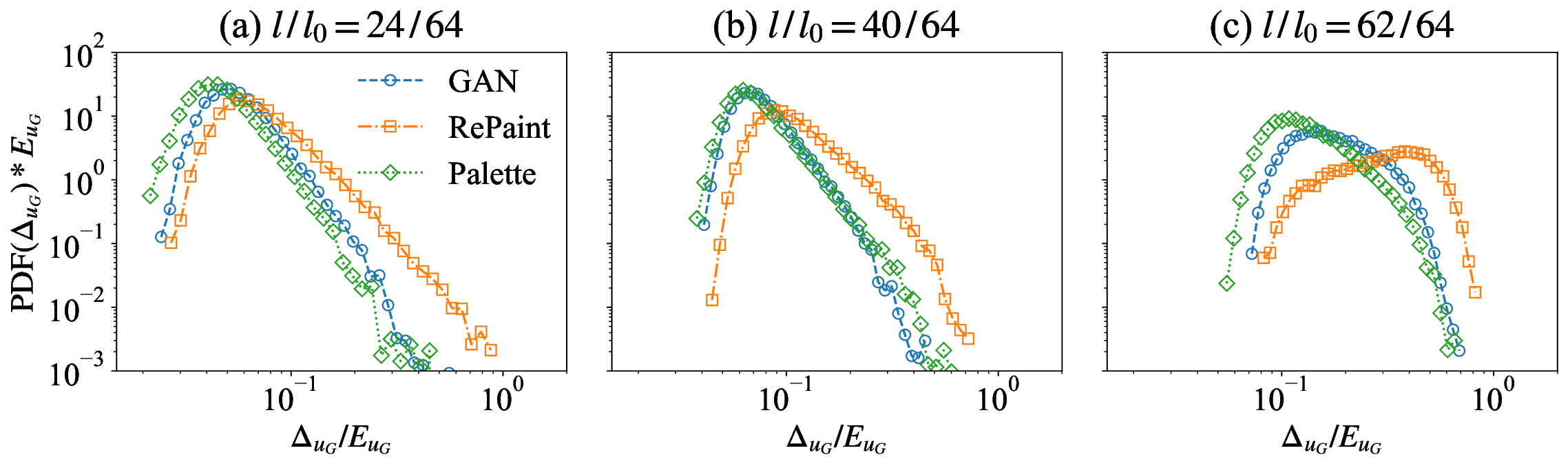}
\caption{The PDFs of the spatially averaged $L_2$ error for a single flow configuration obtained from GAN, RePaint and Palette models. The gap size changes from (\textbf{a}) $l/l_0=24/64$, to (\textbf{b}) $40/64$ and (\textbf{c}) $62/64$.\label{fig:PDF_L2_c}}
\end{figure}

In Figure \ref{fig:PDF_L2_c}, we compare the PDFs of the spatially
averaged $L_2$ error, $\Delta_{u_G}$, for different flow
configurations. For small and medium gap sizes (Figure
\ref{fig:PDF_L2_c}\textbf{a},\textbf{b}), the PDFs of GAN and Palette
closely match, whereas the PDF of RePaint, although similar in shape,
exhibits a range with larger errors. For the largest gap size (Figure
\ref{fig:PDF_L2_c}\textbf{c}), Palette is clearly the most accurate,
predicting the smallest errors.
Again, RePaint performs the worst, characterized by a peak at high error values and a broad error range.

\begin{figure}
\begin{tikzpicture}
    \node[anchor=south west,inner sep=0] (image) at (0,0) {\includegraphics[width=1.0\textwidth]{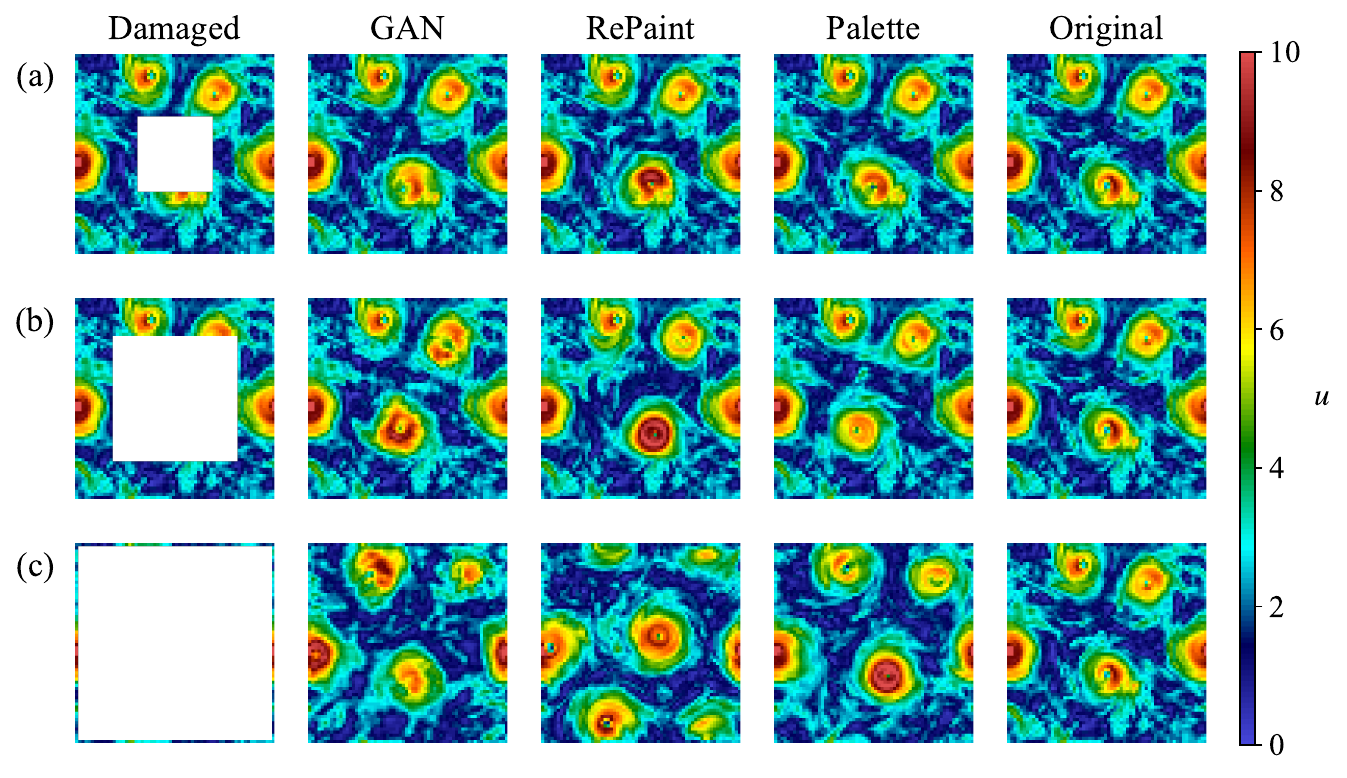}};
    \begin{scope}[x={(image.south east)},y={(image.north west)}]
        \draw[red, thick] (0.217, 0.02) rectangle (0.734, 1);
    \end{scope}
\end{tikzpicture}
\caption{Examples of reconstruction of an instantaneous field (velocity magnitude) for a square gap of size (\textbf{a}) $l/l_0=24/64$, (\textbf{b}) $l/l_0=40/64$ and (\textbf{c}) $l/l_0=62/64$. The damaged fields are shown in the first column, while the second to fourth columns, circled by a red rectangle, show the reconstructed fields obtained from GAN, RePaint and Palette. The ground truth is shown in the fifth column.\label{fig:reconstruction}}
\end{figure}

Finally, Figure \ref{fig:reconstruction} provides a visual qualitative idea of the reconstruction capabilities of the instantaneous velocity magnitude field using the three adopted models. While all methods generally perform well in locating vortex structures within smaller gaps and produce realistic turbulent reconstructions, RePaint is a notable exception. In particular, for the largest gap (in Fig. \ref{fig:reconstruction}\textbf{c}), RePaint's performance lags significantly behind the other two methods, failing to accurately predict vortex positions and resulting in a significantly larger MSE.

\subsection{Multi-scale Information}
\begin{figure}
	\includegraphics[width=1.0\textwidth]{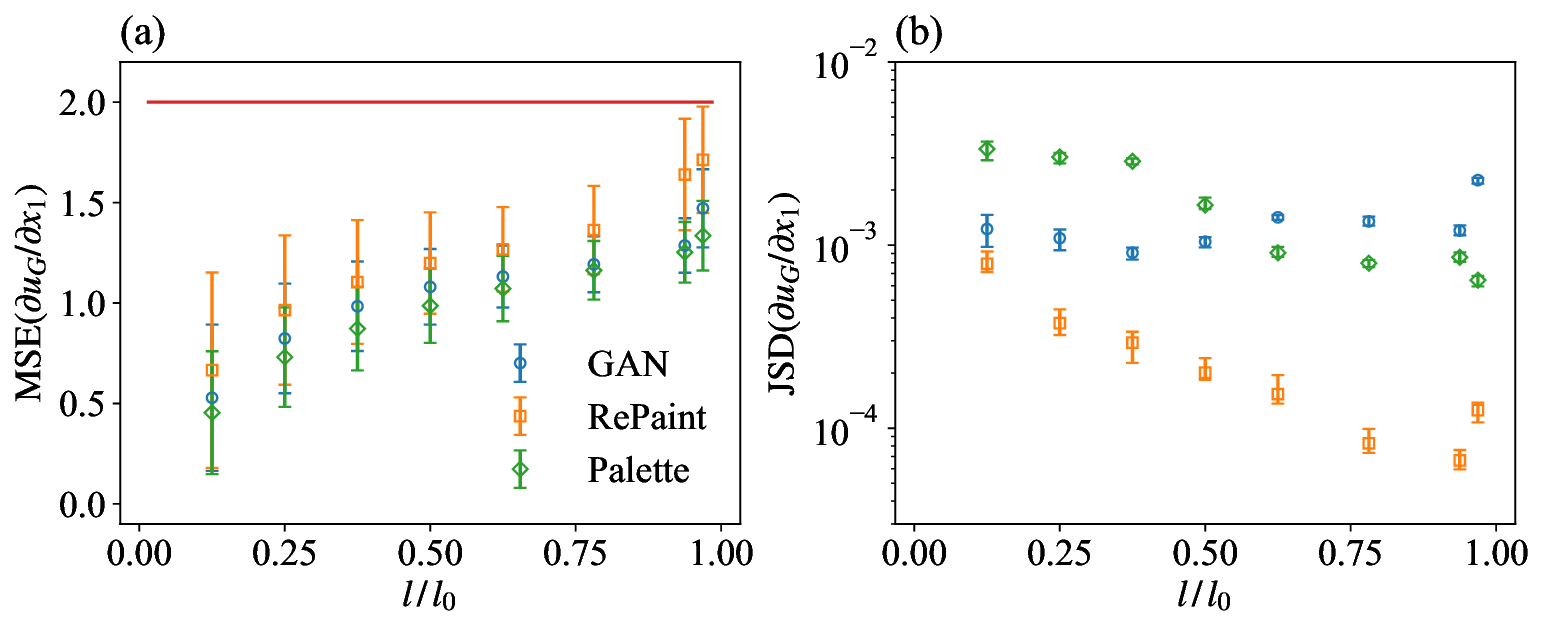}
    \caption{(\textbf{a}) MSE and (\textbf{b}) JS divergence between the PDFs for the gradient of the original and generated velocity magnitude, as obtained from GAN, RePaint and Palette, for a square gap with variable size. The red horizontal line in panel \textbf{a} represents the uncorrelated baseline, equal to $2$. Error bars are obtained in the same way as in Figure \ref{fig:MSE_JSD}.\label{fig:Grad_MSE_JSD}}
\end{figure}

\begin{figure}
    \centering
    \includegraphics[width=\textwidth]{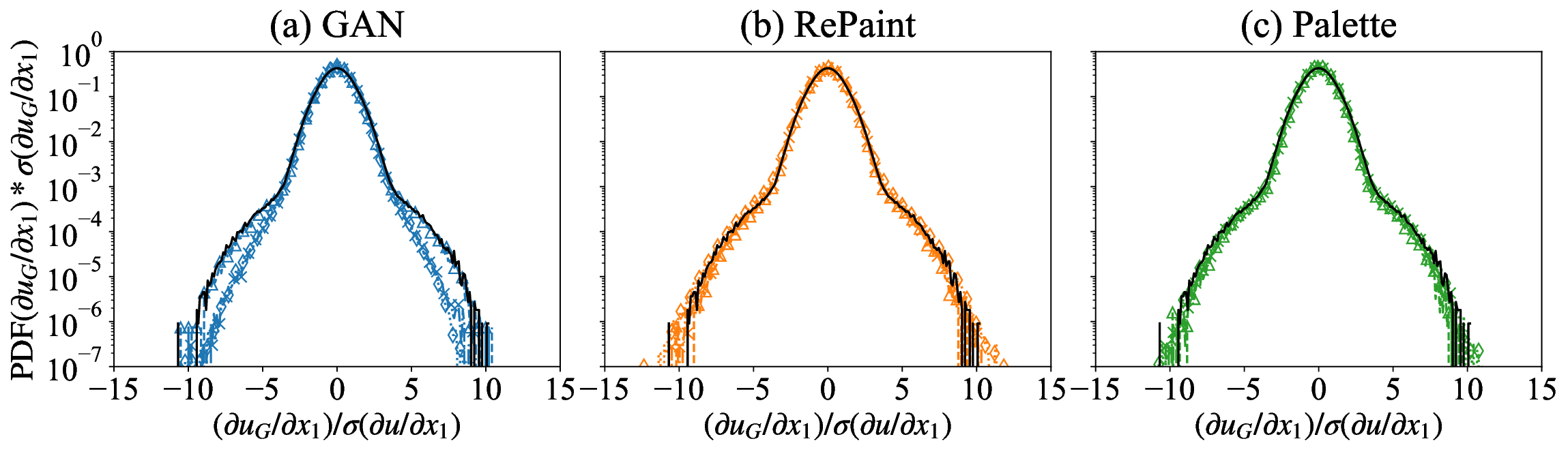}
\caption{The PDFs of the gradient of the reconstructed velocity
  magnitude in the missing region obtained from (\textbf{a}) GAN,
  (\textbf{b}) RePaint and (\textbf{c}) Palette, for a square gap of
  variable size $l/l_0=24/64$ (triangle), $40/64$ (cross), and $62/64$
  (diamond). The PDF of the true data over the whole region is plotted
  for reference (solid black line) and $\sigma(\partial u/\partial
  x_1)$ is the standard deviation of the original data over the full
  domain.\label{fig:Grad_PDF}}
\end{figure}

This section presents a quantitative analysis of the multi-scale information reconstructed by the different methods. We begin by examining the gradient of the reconstructed velocity magnitude in the missing region, denoted as $\partial u_G/\partial x_1$. Figure \ref{fig:Grad_MSE_JSD}\textbf{a} shows the MSE of this gradient, $\mathrm{MSE}(\partial u_G/\partial x_1)$, defined similarly to Equation (\ref{equ:MSE}). The results show that Palette consistently achieves the lowest MSE. GAN's performance is comparable for most gap sizes, but deteriorates significantly at the extremely large gap size. In contrast, while RePaint has larger point-wise reconstruction errors for the gradient, it maintains the smallest JS divergence, $\mathrm{JSD}(\partial u_G/\partial x_1)$, as shown in Figure \ref{fig:Grad_MSE_JSD}\textbf{b}, indicating its robust statistical properties.
For small gap sizes, Palette has larger JS divergence than GAN, while
the situation is reversed at higher gap values (Figure
\ref{fig:Grad_MSE_JSD}\textbf{b}). It is worth noting that, like the
velocity module PDFs, the reconstructed gradient PDFs are very well
matched by all three methods.  In contrast to velocity, the GAN
reconstruction is less accurate for very large gaps in the case of
gradient statistics, as can be seen by comparing the PDFs in the
different panels of Figure~\ref{fig:Grad_PDF}. This last observation
limits the applications of the GAN to the modeling of small-scale
turbulent observables.

\begin{figure}
\begin{tikzpicture}
    \node[anchor=south west,inner sep=0] (image) at (0,0) {\includegraphics[width=1.0\textwidth]{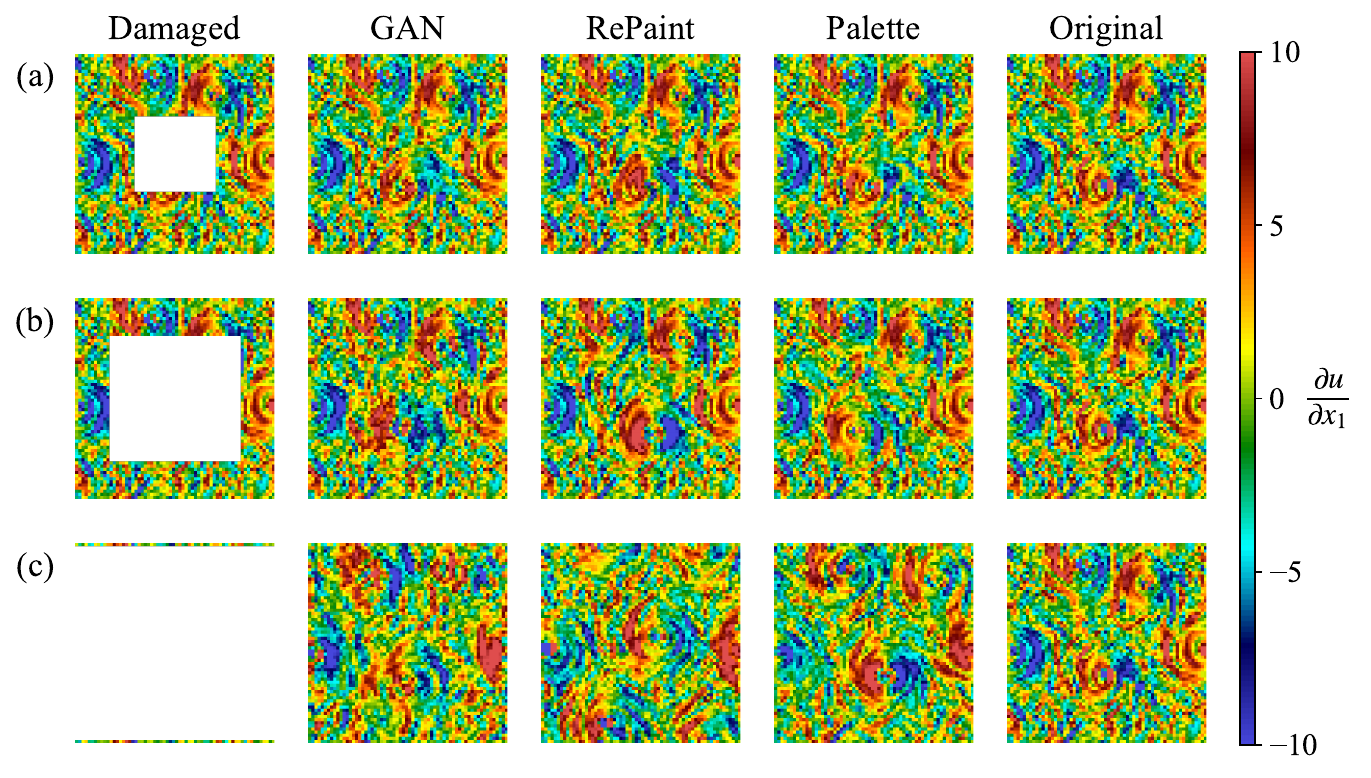}};
    \begin{scope}[x={(image.south east)},y={(image.north west)}]
        \draw[red, thick] (0.213, 0.02) rectangle (0.722, 1);
    \end{scope}
\end{tikzpicture}
\caption{The gradient of the velocity magnitude fields shown in Figure \ref{fig:reconstruction}. The first column shows the damaged fields with a square gap of size (\textbf{a}) $l/l_0=24/64$, (\textbf{b}) $l/l_0=40/64$ and (\textbf{c}) $l/l_0=62/64$. Note that for the case $l/l_0=62/64$, the gap extends almost to the borders, leaving only a single vertical velocity line on both the left and right sides, where the original gradient field is missing. 
The gradient of the reconstructions from GAN, RePaint and Palette, shown in the second to fourth columns, is surrounded by a red rectangle for emphasis, while the fifth column shows the ground truth.}\label{fig:gradient_reconstruction}
\end{figure}

\begin{figure}
    \centering
    \includegraphics[width=\textwidth]{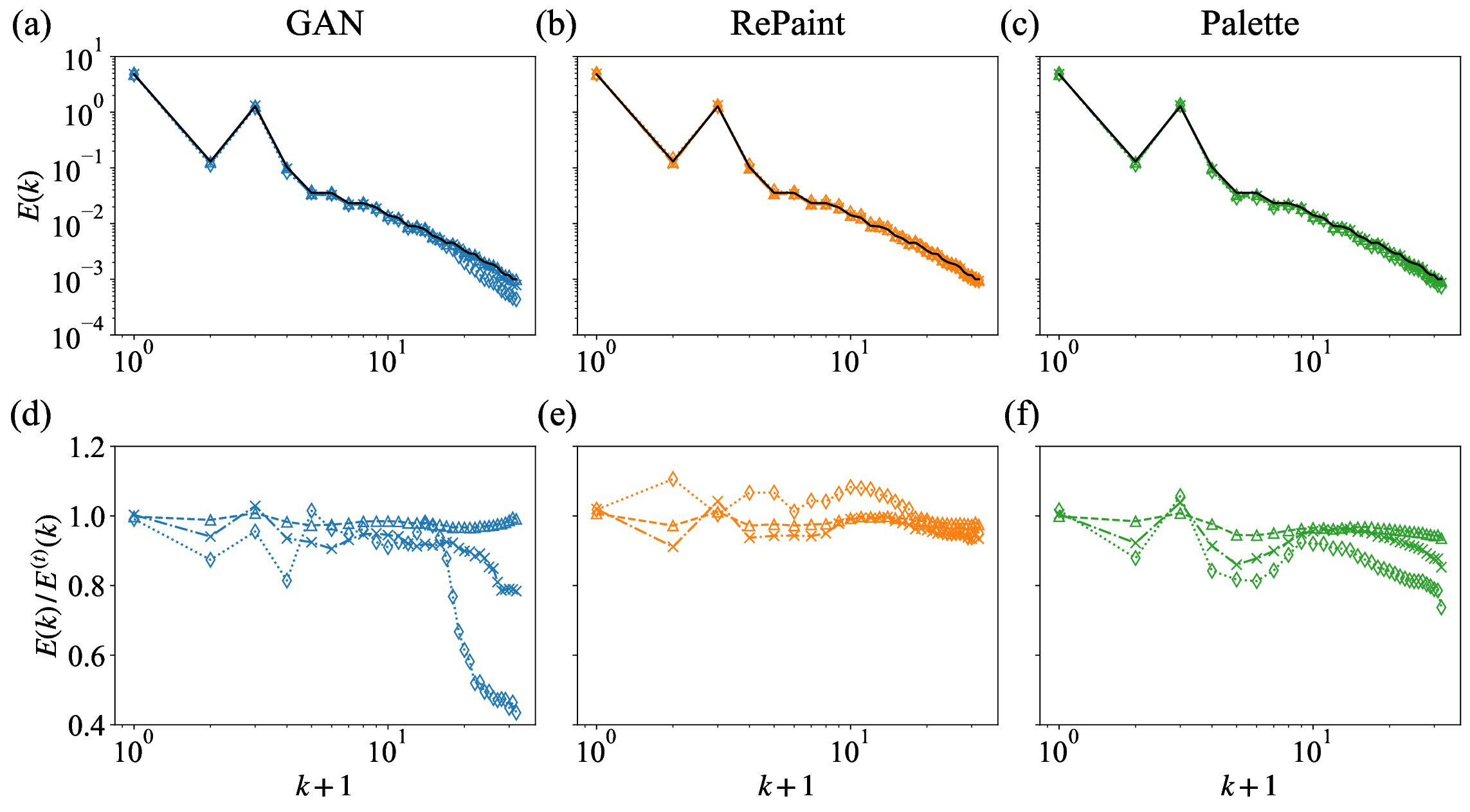}
    \caption{Energy spectra of the original velocity magnitude (solid
      black line) and the reconstructions obtained from (\textbf{a})
      GAN, (\textbf{b}) RePaint and (\textbf{c}) Palette for a square
      gap of sizes $l/l_0=24/64$ (triangle), $40/64$ (cross), and
      $62/64$ (diamond). The corresponding $E(k)/E^{(t)}(k)$ is shown
      in (\textbf{d}--\textbf{f}), where $E(k)$ and $E^{(t)}(k)$ are
      the spectra of the reconstructed fields and the ground truth,
      respectively.\label{fig:spectrum}}
\end{figure}

\begin{figure}
    \centering
    \includegraphics[width=\textwidth]{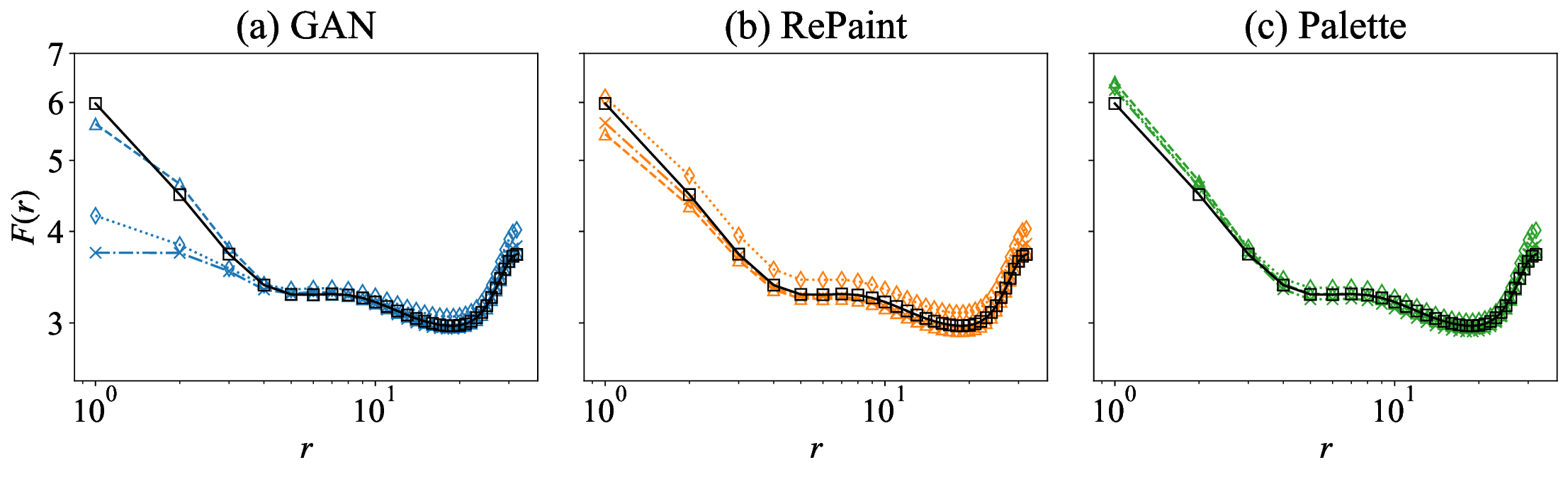}
\caption{The flatness of the original field (solid black line) and the
  reconstructions obtained from (\textbf{a}) GAN, (\textbf{b}) RePaint
  and (\textbf{c}) Palette for a square gap of sizes $l/l_0=24/64$
  (triangle), $40/64$ (cross), and $62/64$
  (diamond).\label{fig:flatness}}
\end{figure}

The results of these methods can be more directly visualized by examining the gradient of the reconstruction samples, as shown in Figure \ref{fig:gradient_reconstruction}. For small gap sizes (Figure \ref{fig:gradient_reconstruction}\textbf{a}), all three methods produce realistic predictions that correlate well with the original structure. However, for medium and large gap sizes (Figure \ref{fig:gradient_reconstruction}\textbf{b},\textbf{c}), only Palette is able to generate gradient structures that are well correlated with the ground truth.
The better performance of DMs in capturing statistical properties is further demonstrated by a scale-by-scale analysis of the 2D energy spectrum obtained from the reconstructed fields, 
\begin{linenomath}
    \begin{equation}
        E(k)=\sum_{k\le\|\bm{k}\|<k+1}\frac{1}{2}\langle\hat{u}(\bm{k})\hat{u}^\ast(\bm{k})\rangle.
    \end{equation}
\end{linenomath}
Here, $\bm{k}=(k_1,k_2)$ denotes the horizontal wavenumber, $\hat{u}(\bm{k})$ is the Fourier transform of the velocity magnitude, and $\hat{u}^\ast(\bm{k})$ is its complex conjugate. Direct comparison of the spectra are shown in Figure \ref{fig:spectrum}\textbf{a}--\textbf{c}, for three gap sizes. In Figure \ref{fig:spectrum}\textbf{d}--\textbf{f}, we plot the ratio of the reconstructed to the original spectra, denoted as $E(k)/E^{(t)}(k)$. Deviations from unity in this ratio better highlight the wavenumber regions where the reconstruction is less accurate. While all methods produce satisfactory energy spectra, a closer examination of the ratio to the original energy spectrum shows that RePaint and Palette maintain uniformly good correspondence across all scales and for all gap sizes. Conversely, GAN performs well at small gap sizes, but exhibits poorer performance at large wavenumbers for medium and large gap sizes.
Consistent results are observed when examining the flatness of the velocity magnitude increments:
\begin{linenomath}
    \begin{equation}\label{equ:flatness}
        F(r)=\langle(\delta_r u)^4\rangle/\langle(\delta_r u)^2\rangle^2,
    \end{equation}
\end{linenomath}
where $\delta_ru=u(\bm{x}+\bm{r})-u(\bm{x})$ and $\bm{r}=(r,0)$, with $\langle\cdot\rangle$ denoting the average over test data and over $\bm{x}$, for points $\bm{x}$ and $\bm{x}+\bm{r}$ where only one, or both of them, are within the gap. The flatness calculated over the entire region of the original field is also shown for comparison. In Figure \ref{fig:flatness}, the flatness results further confirm that RePaint and Palette consistently maintain their high-quality performance across all scales. In contrast, while GAN is effective at small gap sizes, it faces challenges in maintaining similar standards at small scales for medium and large gap sizes.

\section{Probabilistic Reconstructions with DMs}\label{sec:ProbRecs}
So far, we have analyzed the performances of the three models in the reconstruction of the velocity magnitude itself and its statistical properties. 
In this section, we explore the probabilistic reconstruction capabilities of DMs, i.e. the fact that DMs provide us with many possible reconstructions that we can quantify in terms of a mean error and a variance. This is a significant advantage over the GAN architecture we have used in this work. It is worth noting that the implementation of stochastic GANs is also possible, although out of interest for our analysis.
Focusing on a specific gap size, we select two flow configurations: the first is a configuration for which the discrepancy between the reconstructed fields and the true data--as quantified by the mean $L_2$ error--is small and comparable across GAN, Palette and RePaint; the second is a more complex situation for reconstruction, as all models display large discrepancies to the true data. For each of these two configurations, we performed 20,480 reconstructions using RePaint and Palette.

\begin{figure}
\begin{tikzpicture}
    \node[inner sep=0] (imgA) {
    \begin{minipage}{0.5\textwidth}
        \includegraphics[width=1.0\textwidth]{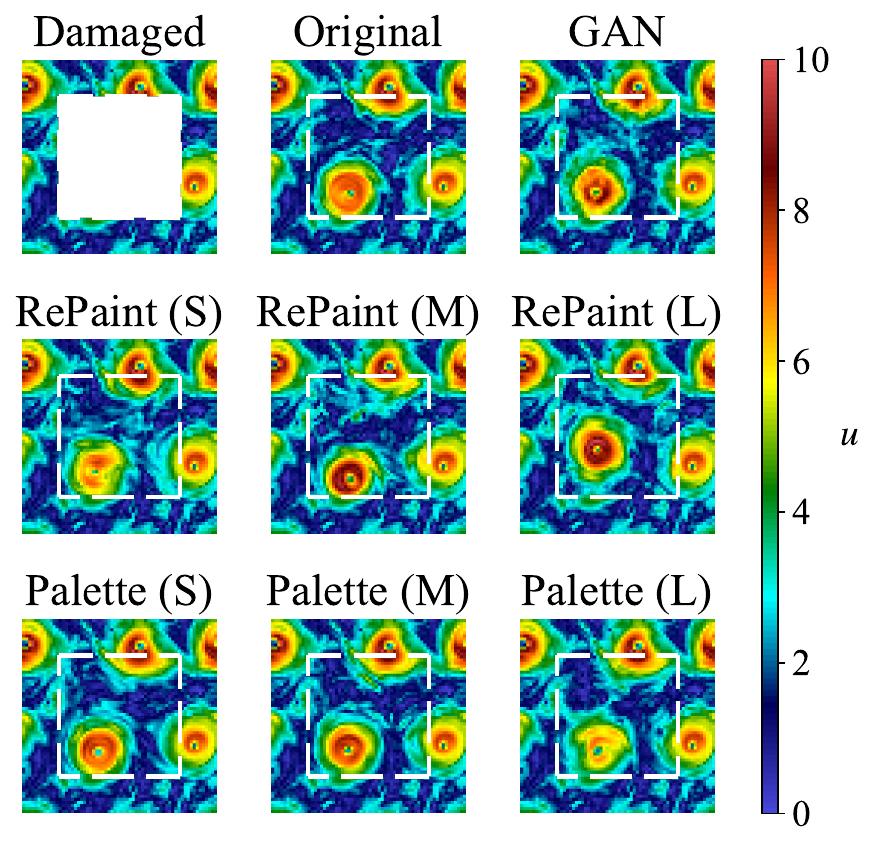}
    \end{minipage}};
    \node[anchor=north west] at ([xshift=-2em, yshift=-2em] imgA.north west) {(b)};
    \node[anchor=north west] at ([xshift=-2em, yshift=-8.8em] imgA.north west) {(c)};
    \node[anchor=north west] at ([xshift=-2em, yshift=-15.6em] imgA.north west) {(d)};
    \node[inner sep=0, left=2em of imgA.north west, anchor=north east] (imgB) {
    \begin{minipage}{0.5\textwidth}
        \includegraphics[width=1.0\textwidth]{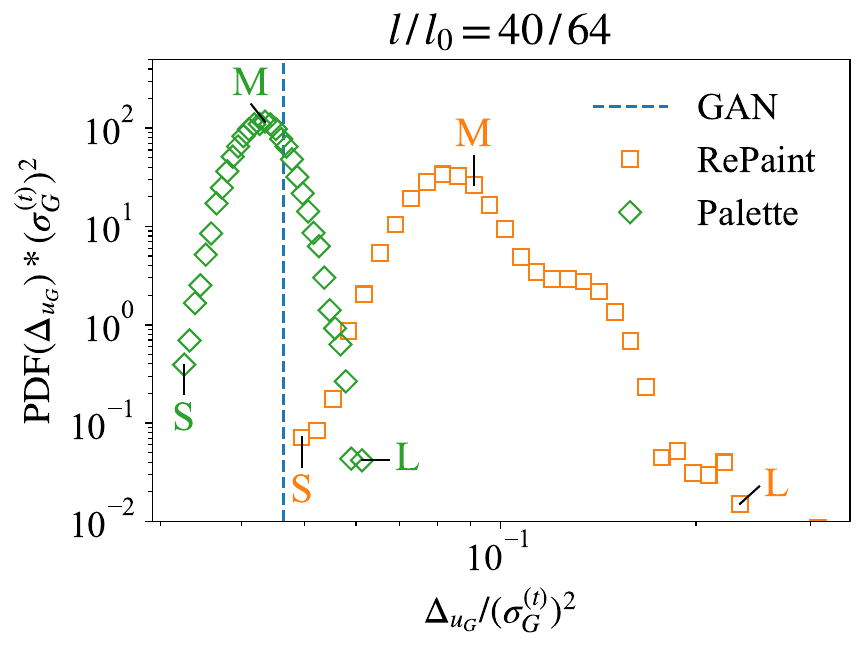}
    \end{minipage}};
    \node[anchor=north west] at (imgB.north west) {(a)};
    \node[draw, rectangle, red, thick, minimum width=0.275\textwidth, minimum height=0.154\textwidth, anchor=north west] at (imgA.north west) {};
\end{tikzpicture}
    \caption{ Probabilistic reconstructions from DMs for a fixed
      measurement outside a square gap with size $l/l_0=40/64$ for a
      configuration where all models give pretty small reconstruction
      errors.  (\textbf{a}) PDFs of the spatially averaged $L_2$ error
      over different reconstructions obtained from RePaint and
      Palette. The blue vertical dashed line indicates the error for
      the GAN case. (\textbf{b}) The damaged measurement and ground
      truth, circled by a red rectangle, and the prediction from
      GAN. (\textbf{c}) The reconstructions from RePaint with a small
      $L_2$ error (S), the mean $L_2$ error (M) and with a large $L_2$
      error (L). (\textbf{d}) The reconstructions from Palette
      corresponding to a small $L_2$ error (S), the mean $L_2$ error
      (M), and a large $L_2$ error (L).\label{fig:ProbRecs-lg40_idx0}}
\end{figure}
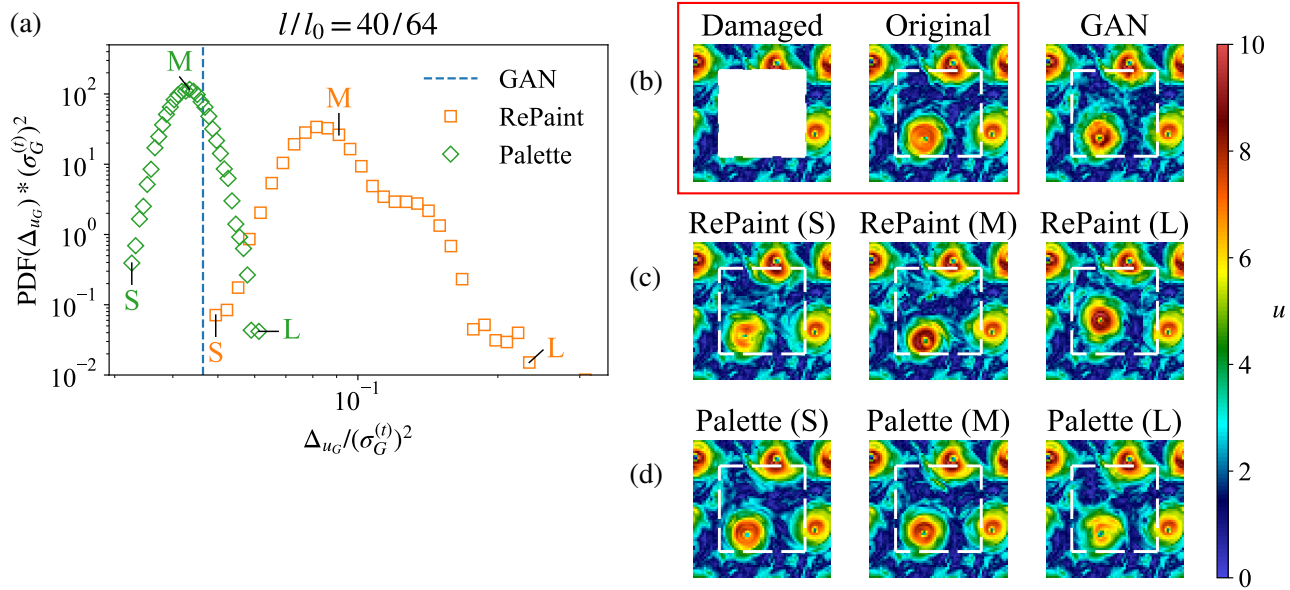

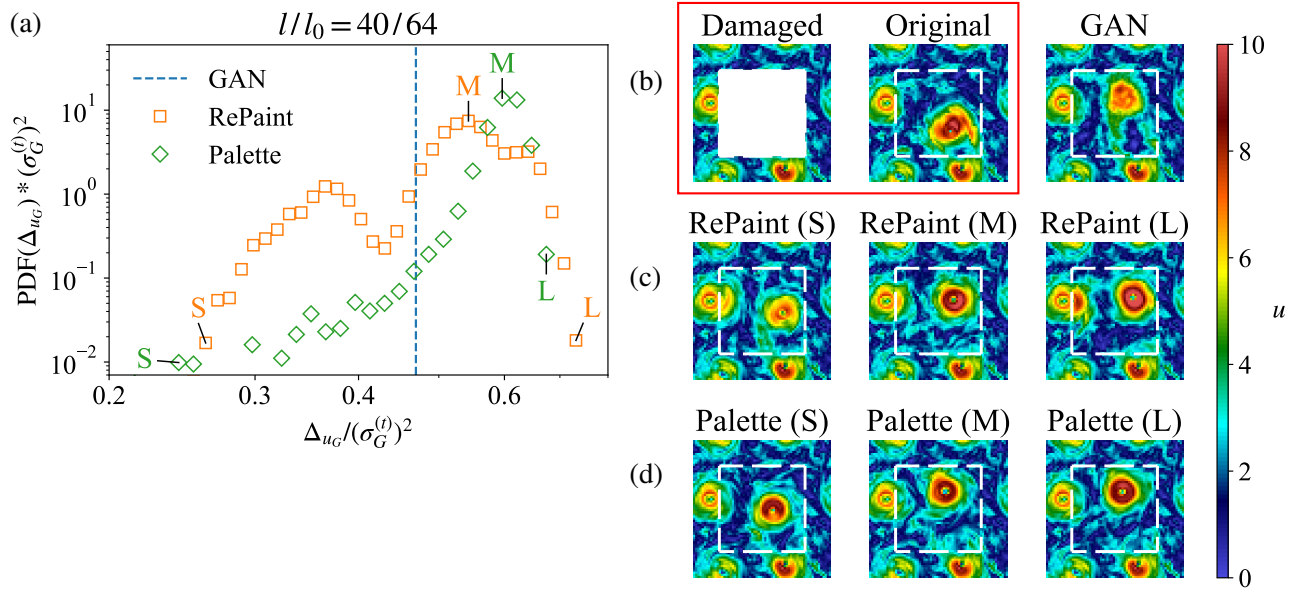
\begin{figure}
\begin{tikzpicture}
    \node[inner sep=0] (imgA) {
    \begin{minipage}{0.5\textwidth}
        \includegraphics[width=1.0\textwidth]{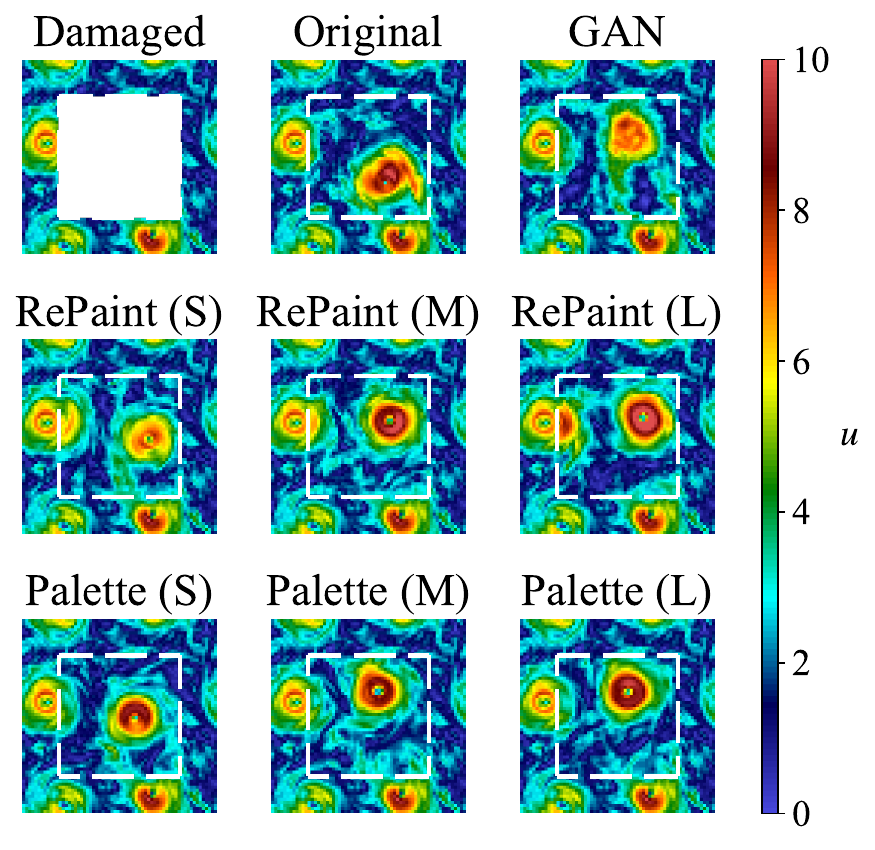}
    \end{minipage}};
    \node[anchor=north west] at ([xshift=-2em, yshift=-2em] imgA.north west) {(b)};
    \node[anchor=north west] at ([xshift=-2em, yshift=-8.8em] imgA.north west) {(c)};
    \node[anchor=north west] at ([xshift=-2em, yshift=-15.6em] imgA.north west) {(d)};
    \node[inner sep=0, left=2em of imgA.north west, anchor=north east] (imgB) {
    \begin{minipage}{0.5\textwidth}
        \includegraphics[width=1.0\textwidth]{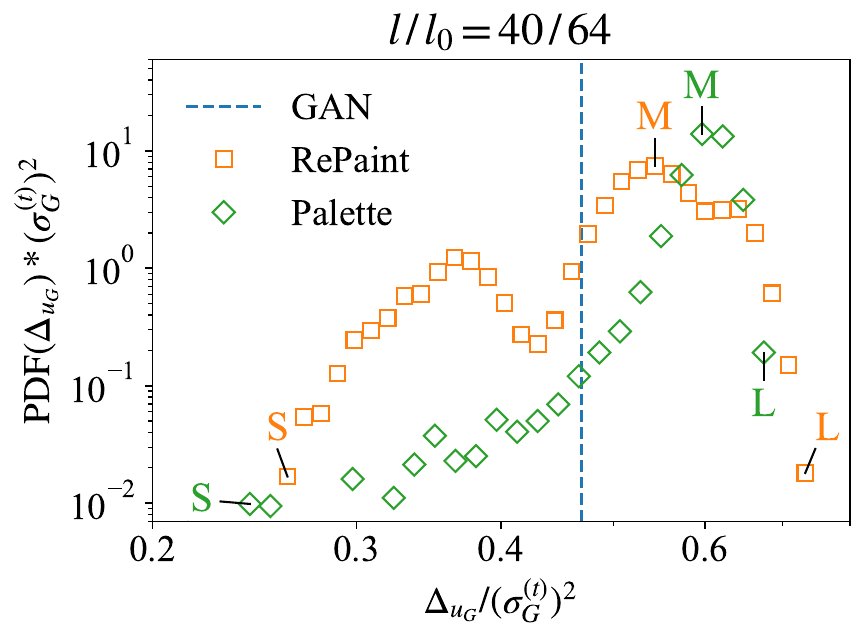}
    \end{minipage}};
    \node[anchor=north west] at (imgB.north west) {(a)};
    \node[draw, rectangle, red, thick, minimum width=0.275\textwidth, minimum height=0.154\textwidth, anchor=north west] at (imgA.north west) {};
\end{tikzpicture}
    \caption{Similar to Figure \ref{fig:ProbRecs-lg40_idx0}, but for a flow configuration chosen for its large reconstruction errors from GAN, RePaint and Palette.\label{fig:ProbRecs-lg40_idx1}}
\end{figure}

Figure \ref{fig:ProbRecs-lg40_idx0}\textbf{a} displays the PDFs of the spatially averaged $L_2$ errors across different reconstruction realizations, compared to GAN's unique reconstruction error indicated by a blue dashed line. The comparison shows that Palette achieves a lower mean $L_2$ error than GAN, along with a smaller variance, indicating high model confidence for this case. In contrast, RePaint tends to produce higher errors with a wider variance. The comparison is more evident in Figure \ref{fig:ProbRecs-lg40_idx0}\textbf{b} where it appears that GAN provides a realistic reconstruction with accurate vortex positioning. As for RePaint, it sometimes inaccurately predicts vortex positions (Figure \ref{fig:ProbRecs-lg40_idx0}\textbf{c} (L)) or fails to accurately represent the energy distribution, even when the position is correct (Figure \ref{fig:ProbRecs-lg40_idx0}\textbf{c} (S) and (M)), leading to larger errors. Conversely, Figure \ref{fig:ProbRecs-lg40_idx0}\textbf{d} shows that Palette consistently predicts the correct position of vortex structures, with variations in vortex shape or energy distribution being the primary factors affecting the narrow reconstruction error distribution.
Figure \ref{fig:ProbRecs-lg40_idx1} presents the same evaluations for
the configuration where all models produce large errors. As shown in
Figure \ref{fig:ProbRecs-lg40_idx1}\textbf{a}, both RePaint and
Palette show significant variance in errors, with their mean errors
exceeding that of GAN. The ground truth, examined in Figure
\ref{fig:ProbRecs-lg40_idx1}\textbf{b}, highlights the inherent
difficulty of this reconstruction scenario. In particular, an entire
vortex structure is missing, and the proximity of two strong vortices
suggests a potential transient state, possibly involving vortex
merging or vortex breakdown. These situations may be rare in the
training data, leading to a complete failure of GAN to accurately
predict the correct vortex position, as shown in Figure
\ref{fig:ProbRecs-lg40_idx1}\textbf{b}. For RePaint, the challenge of
this reconstruction is reflected in the different predictions of
vortex positions. While some of these predictions are more accurate
than GAN's, RePaint also tends to produce incoherence around the gap
boundaries (Figure
\ref{fig:ProbRecs-lg40_idx1}\textbf{c}). Conversely, Figure
\ref{fig:ProbRecs-lg40_idx1}\textbf{d} shows that Palette's
predictions are not only more consistent with the measurements, but
also provide a range of reconstructions with different vortex
positions.

\section{Conclusions and Discussion}
\label{sec:conclusions}
In this study, we investigated the data augmentation ability of DMs
for damaged measurements of 2D snapshots of a 3D rotating turbulence
at moderate Reynolds numbers. The Rossby number is chosen such as to
produce a bidirectional energy cascade to both large and small
scales. Two DM reconstruction methods are investigated: RePaint, which
uses a heuristic strategy to guide an unconditional DM in the flow
generation, and Palette, a conditional DM trained with paired
measurements and missing information. As a benchmark, we compared
these two DMs with the best-performing GAN method on the same data
set.  We showed that there exist a trade-off between obtaining a
reliable $L_2$ error and good statistical reconstruction
properties. Typically, models that are very good for the former are
less accurate for the latter. Overall, according to our analysis,
Palette seems to be the most promising tool considering both
metrics. Indeed, our comparative study shows that while RePaint
consistently exhibits superior statistical reconstruction properties,
it does not achieve small $L_2$ errors. Conversely, Palette achieves
the smallest $L_2$ errors along with very good statistical
results. Moreover, we observe that GAN fails to provide statistical
properties as accurate as the DMs at small scales for medium and large
gaps.

Concerning probabilistic reconstructions, a crucial feature for
turbulent studies and uncertainty quantification for both theoretical
and practical applications, we have evaluated the effectiveness of the
two DM methods on two specific configurations of different
complexity. For the configuration with sufficient information in the
measurement, Palette shows errors that tend to be smaller than GAN and
exhibits a small variance, indicating high model confidence. However,
RePaint faces challenges in accurately predicting large-scale vortex
positions and struggles to achieve accurate energy distribution. This
difficulty partly stems from RePaint's heuristic conditioning
strategy, which cannot effectively guide the generative process using
the measurement. In a more complex scenario characterized by the
presence in the gappy region of an entire large-scale structure, GAN
completely fails to predict the correct vortex position, while both
DMs can localize it with higher precision by taking advantage of
multiple predictions, although RePaint shows incoherence around gap
boundaries.

In summary, this study establishes a new state-of-the-art method for
2D snapshot reconstruction of 3D rotating turbulence using conditional
DMs, surpassing the previous GAN-based approach. The better
performance of DMs over GANs stems from their iterative, denoising
construction process, which builds up the prediction scale-by-scale,
resulting in better performance across all scales. The inherent
stochasticity of this iterative process yields a probabilistic set of
predictions conditioned on the measurement, in contrast to the unique
prediction of the GAN here implemented. Our study opens the way to
further applications for risk assessment of extreme events and in
support of various data assimilation methods. It is important to note
that DMs are significantly more computationally expensive than GANs
due to the iterative inference steps. Despite this, many efforts in
the computer vision field have been devoted to accelerating this
process \cite{song2020denoising, salimans2022progressive}. A promising
avenue for future studies could focus on flows at higher Reynolds
numbers and Rossby numbers, close to the critical transition leading
to the inverse energy cascade, a very complex turbulent scenario where
both 3D and 2D physics coexists in a multi-scale environment.

\vspace{6pt}

\begin{acknowledgments}

This research was funded by European Research Council (ERC) under the
European Union's Horizon 2020 research and innovation programme, grant
agreement number 882340 and by the EHPC-REG-2021R0049 grant. \\ ASL
received partial financial support from ICSC - Centro Nazionale di
Ricerca in High Performance Computing, Big Data and Quantum Computing,
funded by European Union - NextGenerationEU. LB received partial
funding from the program FARE-MUR R2045J8XAW.\\ MB acknowledges the
hospitality of the Eindhoven University of Technology TUe, where part
of this work was carried out.\\

DATA: The 2D snapshots of velocity data from rotating turbulence used
in this study are openly available on the open-access Smart-TURB
portal (\url{http://smart-turb.roma2.infn.it}), under the TURB-Rot
repository \cite{biferale2020turb}. The codebase for the two DM
reconstruction methods, RePaint and Palette, is available at
\url{https://github.com/SmartTURB/repaint-turb} and
\url{https://github.com/SmartTURB/palette-turb}, respectively.
\end{acknowledgments}

\appendix

\section[\appendixname~\thesection]{Training Objective of DM for Flow Field Generation}\label{sec:appA}
A notable property of the forward process is that it allows closed-form sampling of $\CV_I^{(n)}$ at any given diffusion step $n$ \cite{weng2021diffusion}. With definitions of $\alpha_n\coloneqq1-\beta_n$ and $\bar{\alpha}_n\coloneqq\Pi_{i=1}^{n}\alpha_i$, we have
\begin{linenomath}
    \begin{equation}
        q(\CV_I^{(n)}|\CV_I^{(0)})\to\CV_I^{(n)}\sim\mathcal{N}(\sqrt{\bar{\alpha}_n}\CV_I^{(0)},(1-\bar{\alpha}_n)\bm{I}).
    \end{equation}
\end{linenomath}
Specifically, for any real flow field $\CV_I^{(0)}$, we can directly evaluate its state after $n$ diffusion steps using
\begin{linenomath}
    \begin{equation}\label{equ:q_sample}
        \CV_I^{(n)}=\sqrt{\bar{\alpha}_n}\CV_I^{(0)}+\sqrt{1-\bar{\alpha}_n}\epsilon,
    \end{equation}
\end{linenomath}
where $\epsilon\sim\mathcal{N}(\bm{0},\bm{I})$.

To optimize the negative log likelihood, $\mathbb{E}_{q(\CV_I^{(0)})}[-\log(p_\theta(\CV_I^{(0)}))]$, which is numerically intractable, we focus on optimizing its usual variational bound:
\begin{linenomath}
    \begin{equation}
        L\coloneqq\mathbb{E}_{q(\CV_I^{(0)})}\mathbb{E}_{q(\CV_I^{(1:N)}|\CV_I^{(0)})}\left[-\log\frac{p_\theta(\CV_I^{(0:N)})}{q(\CV_I^{(1:N)}|\CV_I^{(0)})}\right]\ge\mathbb{E}_{q(\CV_I^{(0)})}[-\log(p_\theta(\CV_I^{(0)}))].
    \end{equation}
\end{linenomath}
The objective can be further reformulated as a combination of KL divergences, denoted as $\kl{\cdot}{\cdot}$, plus an additional entropy term \cite{sohl2015deep,
ho2020denoising}:
\begin{linenomath}
    \begin{align}
        L=\;&\mathbb{E}_{q(\CV_I^{(0)})}\bigg[\underbrace{\kl{p(\CV_I^{(N)}|\CV_I^{(0)})}{p_{\theta}(\CV_I^{(N)})}}_{L_N}\nonumber\\
        &+\sum_{n>1}^N\underbrace{\kl{p(\CV_I^{(n-1)}|\CV_I^{(n)},\CV_I^{(0)})}{p_\theta(\CV_I^{(n-1)}|\CV_I^{(n)})}}_{L_{n-1}}\underbrace{-\log p_\theta(\CV_I^{(0)}|\CV_I^{(1)})}_{L_0}\bigg]
    \end{align}
\end{linenomath}
The first term, $L_N$, has no learnable parameters as $p_\theta(\CV_I^{(N)})$ is a Gaussian distribution, and can therefore be ignored during training. The terms within the second part of the summation, $L_{n-1}$, represent the KL divergence between $p_\theta(\CV_I^{(n-1)}|\CV_I^{(n)})$ and the posteriors of the forward process conditioned on $\CV_I^{(0)}$, which are tractable using Bayes' theorem \cite{weng2021diffusion, li2023synthetic}:
\begin{linenomath}
    \begin{equation}
        p(\CV_I^{(n-1)}|\CV_I^{(n)},\CV_I^{(0)})\to\CV_I^{(n-1)}\sim\mathcal{N}(\tilde{\mu}(\CV_I^{(n)},\CV_I^{(0)}),\tilde{\beta}_n\bm{I}),
    \end{equation}
\end{linenomath}
where
\begin{linenomath}
    \begin{equation}\label{equ:posterior_mean}
        \tilde{\mu}_n(\CV_I^{(n)},\CV_I^{(0)})\coloneqq\frac{\sqrt{\bar{\alpha}_{n-1}}\beta_n}{1-\bar{\alpha}_n}\CV_0+\frac{\sqrt{\alpha_n}(1-\bar{\alpha}_{n-1})}{1-\bar{\alpha}_n}\CV_I^{(n)} 
    \end{equation}
\end{linenomath}
and
\begin{linenomath}
    \begin{equation}
        \tilde{\beta}_n\coloneqq\frac{1-\bar{\alpha}_{n-1}}{1-\bar{\alpha}_n}\beta_n.
    \end{equation}
\end{linenomath}
By setting $\Sigma_\theta=\sigma_n^2\bm{I}$ to untrained constants, where $\sigma_n^2$ can be either $\beta_n$ or $\tilde{\beta}_n$ as discussed in \cite{ho2020denoising}, the KL divergence between the two Gaussians in Equations (\ref{equ:forward}) and (\ref{equ:backward}) can be expressed as
\begin{linenomath}
    \begin{equation}\label{equ:Lnm1_mu}
        L_{n-1}=\mathbb{E}_{q(\CV_I^{(0)})}\left[\frac{1}{2\sigma_n^2}\|\tilde{\mu}_n(\CV_I^{(n)},\CV_I^{(0)})-\mu_\theta(\CV_I^{(n)},n)\|^2\right].
    \end{equation}
\end{linenomath}
Given the Gaussian form of $p_\theta(\CV_I^{(0)}|\CV_I^{(1)})$ as presented in Equation (\ref{equ:backward}), the term $L_0$ also results in the same form as Equation (\ref{equ:Lnm1_mu}). Substituting Equation (\ref{equ:q_sample}) into Equation (\ref{equ:posterior_mean}), we can express the mean of the conditioned posteriors as
\begin{linenomath}
    \begin{equation}
        \tilde{\mu}(\CV_I^{(n)},\CV_I^{(0)})=\frac{1}{\sqrt{\alpha_n}}\left(\CV_I^{(n)}-\frac{\beta_n}{\sqrt{1-\bar{\alpha}_n}}\epsilon\right).
    \end{equation}
\end{linenomath}
Given that $\CV_I^{(n)}$ is available as input to the model, the parameterization can be chosen as
\begin{linenomath}
    \begin{equation}
        \mu_\theta(\CV_I^{(n)},n)=\frac{1}{\sqrt{\alpha_n}}\left(\CV_I^{(n)}-\frac{\beta_n}{\sqrt{1-\bar{\alpha}_n}}\epsilon_\theta(\CV_I^{(n)},n)\right),
    \end{equation}
\end{linenomath}
where $\epsilon_\theta$ is the predicted cumulative noise added to the current intermediate $\CV_I^{(n)}$. This re-parameterization simplifies Equation (\ref{equ:Lnm1_mu}) as
\begin{linenomath}
    \begin{equation}
        L_{n-1}=\mathbb{E}_{q(\CV_I^{(0)}),\,\epsilon}\left[\frac{\beta_n^2}{2\sigma_n^2\alpha_n(1-\bar{\alpha}_n)}\|\epsilon-\epsilon_\theta\left(\CV_I^{(n)}(\CV_I^{(0)},\epsilon),n\right)\|^2\right].
    \end{equation}
\end{linenomath}
In practice, we ignore the weighting term and optimize the following simplified variant of the variational bound:
\begin{linenomath}
    \begin{equation}\label{equ:L_simple}
        L_\mathrm{simple}=\mathbb{E}_{n,\,q(\CV_I^{(0)}),\,\epsilon}\left[\|\epsilon-\epsilon_\theta\left(\CV_I^{(n)}(\CV_I^{(0)},\epsilon),n\right)\|^2\right],
    \end{equation}
\end{linenomath}
where $n$ is uniformly distributed between $1$ and $N$. As demonstrated in \cite{ho2020denoising}, this approach improves sample quality and simplifies implementation.

\section[\appendixname~\thesection]{Implementation Details of DMs for Flow Field Reconstruction}\label{sec:appB}

During the training of both RePaint and Palette models, we set the
total number of diffusion steps $N=2000$. A linear variance schedule
is used, where the variances increase linearly from $\beta_1=10^{-6}$
to $\beta_N=0.01$. Each model employs a U-Net architecture
\cite{ronneberger2015u} characterized by two primary components: a
downsampling stack and an upsampling stack, as shown in Figure
\ref{fig:RePaint}\textbf{a} and Figure
\ref{fig:Palette}\textbf{a}. The configuration of the upsampling stack
mirrors that of the downsampling stack, creating a symmetrical
structure. Each stack performs four steps of downsampling or
upsampling, respectively. These steps consist of several residual
blocks, some steps also include attention blocks. The two stacks are
connected by an intermediate module, which consists of two residual
blocks sandwiching an attention block
\cite{vaswani2017attention}. Both DMs are trained with a batch size of
$256$ on four NVIDIA A100 GPUs for approximately 24 hours.

For the RePaint model, the U-Net stages from the highest to lowest
resolution ($64\times64$ to $8\times8$) are configured with
$[C,2C,3C,4C]$ channels, where $C$ equals $128$. Three residual blocks
are used at each stage. Attention mechanisms, specifically multi-head
attention with four heads, are implemented after each residual block
at the $16\times16$ and $8\times8$ resolution stages, and also within
the intermediate module (Figure \ref{fig:RePaint}\textbf{a}). The
model is trained using the AdamW optimizer
\cite{loshchilov2017decoupled} with a learning rate of $10^{-4}$ over
$2\times10^5$ iterations. In addition, an exponential moving average
(EMA) strategy with a decay rate of $0.999$ is applied over the model
parameters. During the reconstruction phase with a total of $N=2000$
diffusion steps, the resampling technique is initiated at $n=990$ and
continues down to $n=0$. In this approach, resampling is applied at
every 10th step within this range, resulting in its application at
$100$ different points. At each point the resampling involves a jump
size of $j=10$ and this procedure is iterated $9$ times for each
resampling point.

For the Palette model, the U-Net configuration uses $[C, 2C, 4C, 8C]$
channels across its stages, with $C$ set to $64$. Each stage has two
residual blocks. Attention mechanisms are uniquely implemented in the
intermediate module, with multi-head attention using $32$ channels per
head, as shown in Figure \ref{fig:Palette}\textbf{b}. The model also
incorporates a dropout rate of $0.2$ for regularization. Following the
approach in \cite{chen2020wavegrad, saharia2022image,
  saharia2022palette}, we train Palette by conditioning the model on
the continuous noise level $\bar{\alpha}$, instead of the discrete
step index $n$. As a result, the loss function originally formulated
in Equation (\ref{equ:L_simple}) is modified to
\begin{linenomath}
    \begin{equation}
        L_\mathrm{simple}=\mathbb{E}_{\bar{\alpha},\,q(\CV_S,\CV_G^{(0)}),\,\epsilon}\left[\|\epsilon-\epsilon_\theta\left(\CV_S,\CV_G^{(n)}(\CV_G^{(0)},\epsilon),\bar{\alpha}\right)\|^2\right].
    \end{equation}
\end{linenomath}
In this process, we first uniformly sample $n$ from $1$ to $N$, and
then uniformly sample $\bar{\alpha}$ in the range from
$\bar{\alpha}_{n-1}$ to $\bar{\alpha}_{n}$. This approach allows
Palette to use different noise schedules and total backward steps
during inference. In fact, during reconstruction we use a total of
$1000$ backward steps with a linear noise schedule ranging from
$\beta_1=10^{-4}$ to $\beta_N=0.09$. The Adam optimizer
\cite{kingma2014adam} is used with a learning rate of
$5\times10^{-5}$, training the model for approximately $720$ to $750$
epochs.


\bibliography{references}

\end{document}